\documentclass{sig-alternate}
\usepackage{mathptmx} 

\newcommand{\ignore}[1]{}
\usepackage[pass]{geometry}
\usepackage{fancyhdr}
\usepackage[normalem]{ulem}
\usepackage[hyphens]{url}
\usepackage{hyperref}
\usepackage{color}
\usepackage{soul}

\usepackage{microtype} 

\fancypagestyle{firstpage}{
	\fancyhf{}
	\setlength{\headheight}{50pt}
	
	\pagenumbering{arabic}
}

\usepackage{amssymb,amsmath,amsfonts}
\usepackage{graphicx}
\usepackage{subfig}
\usepackage{algorithm}
\usepackage{algorithmic}
\newcommand{\redHL}[1]{\textcolor{red}{#1}}

\newcommand{\shrink}{Eliminate vertical white-space}
\newcommand{\vshrink}[1]{
  \ifdefined\shrink
    \vspace{-#1cm}
  \else
    \vspace{0cm}
  \fi
}

\newcommand{\arch}{CRAM-PM}

\title{Computational RAM to Accelerate String Matching at Scale}

\numberofauthors{9}
\author{
	\alignauthor Zamshed Iqbal Chowdhury\\
	\texttt{chowh005@umn.edu}\\
	\texttt{University of Minnesota}
	\and
	\alignauthor S. Karen Khatamifard \\
	\texttt{khatami@umn.edu}\\
	\texttt{University of Minnesota}
	\and
	\alignauthor Zhengyang Zhao\\
	\texttt{zhaox526@umn.edu}\\
	\texttt{University of Minnesota}
	\and
	\alignauthor Masoud Zabihi\\
	\texttt{zabih003@umn.edu}\\
	\texttt{University of Minnesota}
	\and
	\alignauthor Salonik Resch\\
	\texttt{resc0059@umn.edu}\\
	\texttt{University of Minnesota}
	\and
	\alignauthor Meisam Razaviyayn\\
	\texttt{razaviya@usc.edu}\\
	\texttt{University of Southern California}
	\and
	\alignauthor Jian-Ping Wang\\
	\texttt{jpwang@umn.edu}\\
	\texttt{University of Minnesota}
	\and
	\alignauthor Sachin S. Sapatnekar\\
	\texttt{sachin@umn.edu}\\
	\texttt{University of Minnesota}
	\and
	\alignauthor Ulya R. Karpuzcu\\
	\texttt{ukarpuzc@umn.edu}\\
	\texttt{University of Minnesota}
}
\ignore{
\numberofauthors{9}
\author{
\alignauthor Zamshed I. Chowdhury\\
\affaddr{
	University of Minnesota
	}\\
\affaddr{
	Minneapolis, MN
	}\\
\email{chowh005@umn.edu}
\alignauthor
S. Karen Kahtamifard\\
\affaddr{University of Minnesota}\\
\affaddr{Minneapolis, MN}\\
\email{khatami@umn.edu}
\alignauthor
Zhengyang Zhao\\
\affaddr{University of Minnesota}\\
\affaddr{Minneapolis, MN}\\
\email{zabih003@umn.edu}
\alignauthor
Masoud Zabihi\\
\affaddr{University of Minnesota}\\
\affaddr{Minneapolis, MN}\\
\email{chowh005@umn.edu}
\alignauthor
Salonik Resch\\
\affaddr{University of Minnesota}\\
\affaddr{Minneapolis, MN}\\
\email{resc0059@umn.edu}
\alignauthor
Meisam Razaviyayn\\
\affaddr{University of Southern California}\\
\affaddr{Los Angeles, CA}\\
\email{razaviya@usc.edu}
\alignauthor
Jian-Ping Wang\\
\affaddr{University of Minnesota}\\
\affaddr{Minneapolis, MN}\\
\email{jpwang@umn.edu}
\alignauthor
Sachin Sapatnekar\\
\affaddr{University of Minnesota}\\
\affaddr{Minneapolis, MN}\\
\email{sachin@umn.edu}
\alignauthor
Ulya R. Karpuzcu\\
\affaddr{University of Minnesota}\\
\affaddr{Minneapolis, MN}\\
\email{ukarpuzcu@umn.edu}
}
}
\begin{document}
	\maketitle
	\thispagestyle{firstpage}
	\pagestyle{plain}


\noindent
\begin{abstract}
	Traditional Von Neumann computing is falling apart in the era of exploding data
volumes as the overhead of data transfer becomes forbidding. Instead, it is more
energy-efficient to fuse compute capability with memory where the data reside.
This is particularly critical for pattern matching, a key computational step in
large-scale data analytics, which involves repetitive search over very large
databases residing in memory. Emerging spintronic technologies show remarkable
versatility for the tight integration of logic and memory. In this paper, we
introduce \arch, a novel high-density, reconfigurable spintronic in-memory
compute substrate for pattern matching.


\ignore{ The evolution in big data analytics domain presents grand challenges in
computing paradigms,towards achieving energy efficient data processing.
Processing-in-memory has the potential of overcoming the hurdle towards such
goal, however the lack of true in-memory processing substrates still remains the
core challenge. In this paper, we present a spintronics device based
processing-in-memory solution, Computational RAM (CRAM), which can address the
energy efficient issues imposed by big data analytics applications. CRAM is
capable of performing logic operations on data stored in the different rows in
the array, without moving the data, in non-destructive manner. CRAM architecture
is demonstrated through example logic implementation through CRAM array.}

\end{abstract}

\section{Introduction}
\label{sec:introduction}
\noindent Classical computing platforms are not optimized for efficient data transfer,
which complicates large-scale data analytics in the presence of exponentially
growing data volumes. Imbalanced technology scaling further exacerbates this
situation by rendering data communication, and not computation, a critical
bottleneck~\cite{Horow}. Specialization in hardware cannot help in this case unless conducted
in a data-centric manner.  

Tight integration of compute capability into the memory, {\em Processing in memory (PIM)}, is especially
promising as the overhead of data transfer becomes forbidding at
scale.
The rich design space for PIM
spans full-fledged
processors
and co-processors residing in
memory~\cite{nmpTax}.
Until the emergence of 3D-stacking, however, the
incompatibility of the state-of-the-art logic and memory technologies prevented
practical prototype designs. 
Still, 3D-stacking can only achieve {\em processing near memory,
PNM}~\cite{hmc,hbm,amc}.
The main challenge remains to be fusing compute and memory without violating
array regularity.

Emerging spintronic technologies show remarkable versatility for the tight integration of logic and memory.
This paper introduces a {high-density, reconfigurable spintronic}
in-memory compute substrate for pattern matching, \arch, which
fuses compute and memory 
by adding an extra transistor to the standard magnetic tunnel
junction (MTJ) based memory cell~\cite{Lyle10,Wang05}. Thereby each memory cell
can participate in gate-level computation as an input or as an output.
Computation is not disruptive, i.e., memory cells acting as gate inputs do not
loose their stored values. 

\arch\ can implement different types of basic Boolean gates to form a
functionally complete set, therefore there is no fundamental limit to the types
of computation that the array can perform.  Each row can have only one active
gate at a time, however, computation in all rows can proceed in parallel.   
%
\arch\ provides {\em true} in-memory computing by reconfiguring cells within the
memory array to implement logic functions. As all cells in the array are
identical, inputs and outputs to logic gates do not need to be confined to a
specific physical location in the array. 
In other words, \arch\ can intiate computation at any location in the memory
array.

Pattern matching is at the core of many important large-scale data analytics
applications, ranging from bioinformatics to cryptography. The most prevalent
form is string matching via repetitive search over very large
reference databases residing in memory. Therefore, compute substrates such as
\arch, that
collocate logic and memory to 
prevent slow and energy-hungry
data transfers at scale, have great potential. 

In this case, each
step of computation attempts to map a short character string to (the most
similar substring of) an orders-of-magnitude-longer character string, and
repeats this process for a very large number of short strings, where the longer
string is fixed and acts as a reference.

In the following, we analyze a proof-of-concept \arch\ array for large-scale string
matching.
Specifically, 
Section~\ref{sec:cram} covers the basics of how \arch\ fuses compute with
memory; 
Section~\ref{sec:patt} introduces a 
\arch\ implementation for
pattern (string) matching; 
Sections~\ref{sec:eval_setup},~\ref{sec:eval} provide the evaluation;
Section~\ref{sec:rel} compares and contrasts \arch\ to related works;
and Section~\ref{sec:conc} concludes the paper.

\ignore{
Modern processors are inadequately equipped to address the computational
demand of big data analytics as data set sizes grow exponentially with time.
Hardware paradigms have moved towards greater specialization to handle
this challenge, and specialized units for memory-centric computing are 
vital to any future solution.
%
Technology scaling has further enhanced the need for memory-centric computing
as it has improved logic efficiency 
more than data communication. As a result, communication energy dominates computation
energy and 
\ignore{~\cite{Kogge,Horow}.  Table~\ref{tbl:communication_trend}, adapted
from~\cite{parTrend}, compares the cost of computation (a double-precision
fused multiply add) with communication (a 64-bit read from an on-chip SRAM).
For two technology generations: 40nm and 10nm optimized for high performance
(HP), and low power (LP), the communication energy increases from 1.55$\times$
(computation energy) at 40nm to approximately 6$\times$ at 10nm.  Even worse,
transferring the same 64-bit data off-chip, to main memory, would require
more than 50$\times$ computation energy even at 40nm~\cite{parTrend}.  Such
off-chip accesses become increasingly necessary as data sets grow larger, and}
even the cleverest latency-hiding techniques cannot conceal 
the overhead of communication.

An effective way to overcome this bottleneck and maintain data locality
is to embed compute capability into the main memory:
{\em Processing in-memory (PIM)} can effectively address the communication
bottleneck
through distributed processing of data at the
source, obviating the need for intensive energy-hungry communication.  PIM
features a rich design space, which spans full-fledged
processors
and co-processors residing in
memory~\cite{nmpTax}.
However, until recently, such
promising studies could not render practical prototype designs due to the
incompatibility of the state-of-the-art logic and memory technologies.  The
emergence of 3D-stacked architectures solved this problem
partially by enabling {\em processing near-memory,
PNM}~\cite{hmc,hbm,amc},
but genuine {\em processing in-memory} has
remained elusive.

This paper introduces a {\em high-density reconfigurable spintronics-based}
platform providing {\em true processing-in-memory} semantics, as opposed to most
CMOS-based solutions which 
only deliver {processing-near-memory}, leaving the highly demanded potential
in energy-efficiency untapped. 
The resulting {\em Computational RAM}, CRAM, platform
can configure logic blocks of different functionality within a RAM array with
rows which can be accessed in parallel.  Thus, a CRAM-based solution not only
meets true PIM semantics, but also facilitates reconfigurability which
enables tailoring computational and memory resources to the demands of different
algorithms or working sets.

A CRAM array can serve as both, a stand-alone full-fledged processor or a
domain-specific co-processor attached to a host processor.  
From an application standpoint, algorithms are evolving, and such evolution may
render very different requirements for hardware acceleration than
imposed by the previous generation algorithms the hardware is tailored to.
The reconfigurability of CRAM helps it sidestep 
the inflexibility of any non-reconfigurable 
design.
%
The key characteristic of emerging application domains is data-intensity. 

Our preliminary analysis indicates that \arch\ can outperform CMOS solutions in
energy efficiency with competitive throughput: CRAM can deliver
higher throughput at iso-energy, or lower energy consumption at
iso-throughput. 
The idea of the CRAM has been proposed in a brief description
in~\cite{CRAMpatent15}, but as we show, substantial refinement is
necessary to take this idea to a practical implementation across the system
stack.
}


\ignore{

In recent years, the data science is experiencing an unprecedented surge,
following an exponential growth, in data volume. With this growth, big data
analytics are also causing traction over several domains of computational
sciences. The demand for faster and more efficient computational resources is
now higher than ever. From the perspectives of big data analysis, the current
processor designs are not well optimized to handle such data in efficient
manner. To accommodate this rising domain of computational science, computing
paradigms are shifting more and more toward hardware centric specialized
solutions. However, the existing solutions are still limited by high memory
resource requirements for such analytical tasks. On top of that, the asymmetry
between improvement in efficient communication and computation has made this
problem more inclined towards finding a solution which utilizes less
communication and more computation.                

\textit{Processing-in-Memory} (PIM) has emerged as the one of the most promising
solutions, given the constraints imposed by big data analytics applications. It
extends the idea of a system with a centralized processing element to one with
distributed elements with compute capability of different degrees. The idea of
integrating computing capability with memory substrate has the potential of
addressing the aforementioned issues with limitation of memory-bound data
processing. It performs computation on the stored data \textit{in-place}, while
maintaining data locality thereby effectively eliminating most, if not all, of
the energy hungry and long latency memory-bound communications, essentially
improving performance in terms of both energy and latency of overall data
processing.       

The most intriguing thing about \textit{PIM} probably is the diverse design
space in offer which covers from logic stacked memory substrate to full-fledged
processors ~\cite{nmpTax}. Such promise of \textit{PIM} has attracted
researchers to dive into efficient implementation of such solutions. However,
despite numerous effort, there hasn't been any practical prototyping until
recently, mostly due to the incompatibility between conventional logic and
memory technologies. There has been some notable research in \textit{PIM}
domain, which bypassed this issue by clever incorporation of logic and memory.
For instance, 3D architectures enable logic components to be stacked on top of
memory substrate ~\cite{hmc,hbm,amc}. These solutions, however, do not eliminate
memory transaction, rather reduce the energy and latency of such transaction by
reducing the distance between processing element and memory substrate. In this
context, these are not \textit{true} PIM solutions.             

In this paper, we introduce a novel {\em true processing-in-memory} substrate,
based on Magnetic Tunnel Junction (MTJ) devices, called {\em Computational RAM}
or CRAM. The feature that makes this substrate unique is the ability to process
data \textit{in place}, without incurring any data movement overhead of any
sort. The platform is based on high density spintronics devices in an array
structure which are reconfigurable, in runtime, to perform logic operations on
the data. As compared to the CMOS based \textit{processing-near-memory}
solutions, which are not able to benefit from complete elimination of data
communication,  this technology thrives on the truly \textit{local computation}
of stored data. From an organization perspective, the MTJ devices are connected
in a 2-D array structure and the rows of the array can be executed in parallel.
This feature is responsible for its ability to extract inherent parallelism in
data, contributing to the throughput of the array. Furthermore, the same array
can be used to perform different logic operations at different times during
execution. This feature enables it to process data with more complex logic,
thereby effectively reducing processing burden on the centralized processing
element. To extend the impact of such reconfigurability, CRAM can be configured
to be a standalone processing element or domain-specific hardware to assist a
host processor as well. The effect of such a unique substrate is that it could
address the ever evolving  algorithms in a way to satisfy the wide-ranging and
diverse requirement for hardware based acceleration, unlike CMOS based solutions
which are constrained to fit best to the previous generation algorithms. In a
nutshell, CRAM is able to provide an energy efficient true
\textit{processing-in-memory} solution with high throughput. The key observation
here is that the design space of CRAM based solution is extremely rich. For
instance, the same substrate can be configured to satisfy an energy budget or a
throughput target, depending on how the design is carried out.       

While it is a break through progress to come up with a memory substrate for PIM,
it is not merely enough without identifying the potential application domain
that would be benefited from implementation on such substrate. Intuitively, the
type of application for CRAM would be applications with a very high fraction of
memory transactions. These applications would experience improvement in
performance if the high memory traffic and corresponding computing workloads are
replaced by very low energy and moderately low latency computations in memory.
However, having reduced high memory traffic footprint is not merely enough to
extract the full performance benefit from mapping applications in CRAM. Since
CRAM supports row-parallel operations, it is recommended to select applications
which has high degree of data parallelism. For instance, pattern matching
problem would be a good fit for CRAM mapping due the presence of parallel
matching functions between a reference and input patterns. To bolster our claim
that CRAM is indeed a true PIM substrate, we show the mapping of a
bio-informatics pattern matching application i.e. DNA sequence alignment, to
evaluate the performance and energy benefit over CUP and GPU based baseline
implementations. To provide a fair comparison with other \textit{near-memory}
and \textit{in-memory processing} solutions, we quantitatively analyze CRAM
against other substrates.        

The main contributions of this paper are as follows: 
\begin{enumerate}
	
	\item Although the idea of the CRAM has been proposed in a brief description
	in~\cite{CRAMpatent15}, this work presents a more in-depth discussion on the
	CRAM architecture and how the structure affects the performance benefits
	that can be harnessed from it.  
	
	\item A thorough illustration of how logic functionality if fabricated on
	the CRAM substrate, clarified through detailed example implementations. 
	
	\item An example case study of DNA sequence alignment, mapped to CRAM
	substrate, and rigorous evaluation of such implementation with relative
	speedup as compared to other state-of-the-art implementations.  
	
	\item An insight on the potential application domains which can benefit from
	CRAM-based implementation.   
	
\end{enumerate}

The rest of the paper is organized as follows: section II provides details about
the CRAM architecture and organization. The next section presents a few
fundamental compute element designs. Section IV presents an overview to a case
study which is evaluated in successive sections (V and VI). Potential
applications for this novel technology and how existing works relate to this
technology is discussed in sections VII and VIII, before concluding the paper in
section IX.    

}

\section{Background}
\label{sec:cram}  

\subsection{Fusing Compute and Memory}
\label{sec:fuse}

\begin{figure*}[!tp]
	\centering
	\includegraphics[width=\textwidth]{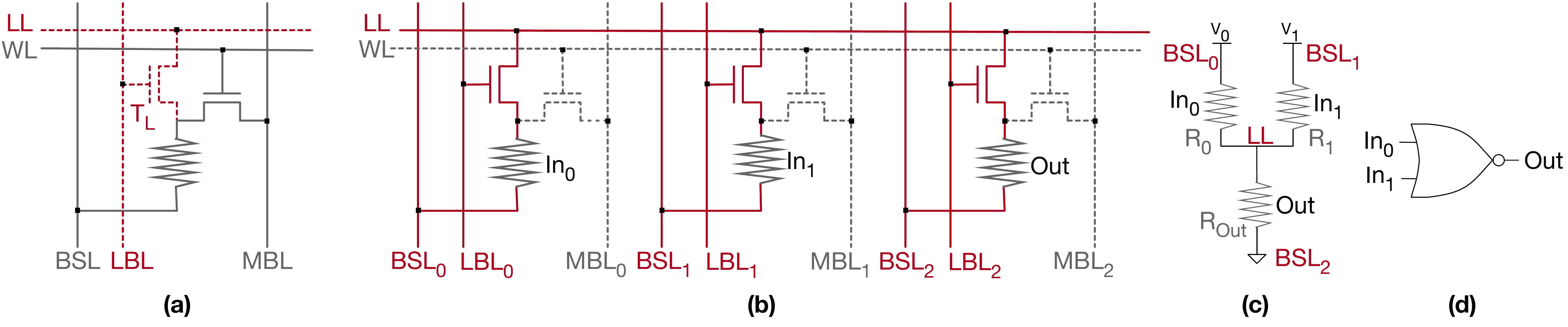}
	\vshrink{0.7}
	\caption{(a) \arch\ cell; (b) 2-input gate formation in the array; 
	  (c), (d) 2-input NOR gate circuit equivalents. \label{fig:basicArray}}
\vshrink{.5}
\end{figure*}

\noindent 
Without loss of generality, we adapt Computational RAM (CRAM)~\cite{cram} as the spintronic PIM substrate to design \arch\ arrays in this study.
In its most basic form, a CRAM array is essentially a 2D magneto-resistive RAM
(MRAM). 
When compared to the standard 1T(ransistor)1M(TJ) MRAM cell, however,
each CRAM cell features an additional transistor $T_L$
(Fig.\ref{fig:basicArray}(a)), which acts as a switch between memory and logic
configurations.
A CRAM cell can operate as a regular MRAM
memory cell or serve as an input/output to a logic gate. 

Each MTJ consists of two layers of ferromagnets, termed as pinned and free
layers, separated by a thin insulator. The magnetic spin
orientation of the pinned layer is fixed; of the free layer, controllable. 
Changing the orientation of the free layer entails
passing a (polarized) current through the MTJ, where the current direction sets
the orientation.  The relative orientation of
the free layer with respect to the pinned layer, i.e., anti-parallel (AP) or
parallel (P), gives rise to two distinct MTJ resistance levels, i.e., 
$R_{high}$ and $R_{low}$, which encode logic 1 and 0, respectively.  As
resistance levels represent logic states, Fig.\ref{fig:basicArray} depicts each
MTJ by its resistance.  

\noindent {\bf Memory Configuration:}
The dashed components in Fig.\ref{fig:basicArray}(a) capture all 
add-ons to the standard MRAM memory cell, in order to support logic functions.
When the array is configured as memory, the {\em Logic Bit Line (LBL)} is set to 0 to turn the switch
$T_L$ off, and thereby to disconnect the cells
from the {\em Logic Line (LL)}.
In this case, the array becomes equivalent to a standard MRAM array. In the following, we detail the configuration for various memory operations (where {\em LBL} is always set to 0).
%
\begin{list}{\labelitemi}{\leftmargin=1em}
\item {Data retention:} The {\em Word Line (WL)} is set to 0 to isolate the cells
  and to prevent current flow through the MTJs.
\item {Read:} {\em WL} is set to 1, to connect each MTJ to its {\em Bit Select Line
  (BSL)} and {\em Memory Bit Line (MBL)}. A small voltage pulse applied between
  {\em BSL} and {\em MBL} induces a current through the MTJ, which is a
  function of the resistance level (i.e., logic state), and which in turn a
  sense amplifier attached to {\em BSL} captures.  
 %
 \item {Write:} {\em WL} is set to 1, to connect each MTJ to its {\em BSL} and {\em MBL}. A large
  enough voltage pulse (in the order of the supply voltage) is applied between
  {\em BSL} and {\em MBL} to induce a large enough current through the MTJ to
  change the spin orientation of the free layer.
\end{list}

\noindent {\bf Logic Configuration:}  {\em LL}
connects all cells participating in computation, on a per row basis. Such cells
may act as logic gate inputs or outputs. 
For each \arch\ cell participating in
computation, {\em WL} is set to 0 to disconnect its MTJ
from {\em MBL}. Instead, {\em LBL} is set to 1 to turn the switch $T_L$ on,
which in turn connects the MTJ to the {\em LL}. 

As an example, Fig.\ref{fig:basicArray}(b) demonstrates the formation of a two input logic gate
in the array, where cells labeled by ``0'', ``1'', and ``2'' correspond to the inputs
$In_0$, $In_1$, and the output $Out$, respectively. 
Fig.\ref{fig:basicArray}(c) depicts the equivalent circuit: {\em BSL} of the
output, $BSL_2$ is grounded, while  {\em BSL} of the two inputs, $BSL_0$ and
$BSL_1$ are set to voltages $V_0$ and $V_1$. The values of $V_0$
and $V_1$ determine the currents through the input MTJs, 
$I_0$ and $I_1$, as a function of their
resistance values $R_0$
and $R_1$ (i.e., logic states). $I_{Out} = I_0 + I_1$ flows
through the output resistance $R_{Out}$. If $I_{Out}$ is higher than the
critical MTJ switching current $I_{crit}$, it will change the free layer
orientation of $Out$'s MTJ, and thereby, the logic state of $Out$. Otherwise, $Out$ will keep
its previous state. 

We can easily expand this example to more than two inputs.
The key observation is that we can change the logic state of the output as a {\em
function} of the logic states of the inputs, within the array.  And voltages at
{\em BSL}s of the inputs dictate how such {\em function}s would look like.

\begin{table}[tp]
\begin{center}
%
\begin{tabular}{cc|c||cc}
  $In_0$ & $In_1$ & $Out$ & $I_{Out} = I_0 + I_1$ &\\
  \hline
  0 ($R_{low}$) & 0 ($R_{low}$) & 1 & $I_{00}$ & $> I_{crit}$ \\
 \hline
 0 ($R_{low}$) & 1 ($R_{high}$) & 0 & $I_{01}$ & $< I_{crit}$ \\
  \hline
  1 ($R_{high}$) & 0 ($R_{low}$) & 0 & $I_{10}= I_{01}$ & $ < I_{crit}$ \\
  \hline
  1 ($R_{high}$) & 1 ($R_{high}$) & 0 & $I_{11}$ & $< I_{crit}$ \\
\end{tabular}
\vshrink{-.1}
\caption{2-input NOR truth table ($Out$ pre-set = 0). \label{tbl:nand}}
\end{center}
\vshrink{1.0}
\end{table}

Continuing with the example from Fig.\ref{fig:basicArray}(b)/(c), let us try to
implement an
universal, 2-input NOR gate. Table~\ref{tbl:nand} provides the truth table. $Out$ would be
0 in this case for all input combinations but $In_0 =0$, $In_1=0$, which incurs
the lowest $R_0$ and $R_1$, and hence, the highest $I_{Out} = I_0 + I_1$. Let us
refer to this value of $I_{Out}$ as $I_{00}$, following Table~\ref{tbl:nand}.
Accordingly, if we pre-set $Out$ to 0 (before computation starts), and 
determine $V_0$ and $V_1$ such that $I_{00}$ does exceed $I_{crit}$, while
both $I_{11}$ and $I_{01}=I_{10}$ does not, $Out$ would not switch
from (its pre-set value) 0 to 1, for all input combinations but $In_0 =0$,
$In_1=0$.

As Boolean gates of practical importance (such as NOR) are commutative, a
single voltage level at the {\em BSL}s of the inputs 
suffices to define a
specific logic functionality. Each voltage level can serve as a signature for a
specific logic gate. Accordingly, in the above example, $V_0 = V_1$ applies, and
its
value simply follows from Kirchoff's Laws, where $R_{high}$, $R_{low}$, and
$I_{crit}$ represent technology dependent constants. In the following, we will
refer to this value as $V_{gate}$. In the example above, $V_{gate} = V_{NOR}$.
While NOR gate is universal, we can implement different types of logic
gates following a similar methodology for mapping the corresponding truth tables
to the \arch\ array. 

\subsection{Basic Computational Blocks}
\label{sec:bb}
\noindent We will next introduce basic \arch\ computational blocks
for pattern matching, including inverters (INV), buffers (COPY), 3-input and
  5-input majority (MAJ) gates, and 1-bit full adders.

\noindent{\bf INV:} INV is a single-input gate. Still, we can follow a similar
methodology to the NOR implementation (Table~\ref{tbl:nand}): Pre-set output to
0, and define $V_{INV}$ in a way such that $I_0$ ($I_1$), i.e., the current if
the input is 0 (1), is higher (lower) than  $I_{crit}$ such that the output does
(not) switch from the pre-set 0 to 1. By definition, $I_1 < I_0$ applies, as
$R_1 > R_0$. 

\noindent{\bf COPY:} For 1-bit copy, two back-to-back invocations of INV can suffice. A more
time and energy efficient implementation, however, can perform the same function
in one step as follows:
Pre-set output to
1, and define $V_{COPY}$ in a way such that $I_0$ ($I_1$), i.e., the current if
the input is 0 (1), is higher (lower) than  $I_{crit}$ such that the output does
(not) switch from the pre-set 1 to 0. By definition, $I_1 < I_0$ applies, as
$R_1 > R_0$. 

\noindent{\bf MAJ:}
MAJ gates accept an odd number of inputs, and assign the majority (logic) state across
all inputs to the output. The structure for a 3-input MAJ3 or 5-input MAJ5 gate
is not any different from the circuit structure in
Fig.~\ref{fig:basicArray}(c) except the higher number of inputs. 
As an example, $I_{Out}$ of the MAJ3 gate assumes its highest value for the
000 assignment of the three inputs -- as the MTJ resistances of the three inputs, $R_0$,
$R_1$, and $R_2$, assume their lowest value for 000.
Any input assignment having at least one 1, gives rise to a lower
$I_{Out}$ than $I_{000}$;
and having at least two 1s,
to an even lower $I_{Out}$.
Finally, $I_{Out}$ reaches its minimum for
the input assignment 111, for which the input MTJs assume their highest
resistance.
Accordingly, we can pre-set the output to 1, and define $V_{MAJ3}$ in a way such
that $I_{Out}$ remains higher than $I_{crit}$ if the three inputs have less than
two 1s, such that $Out$ switches from the pre-set 1 to 0, to match the input majority.
We can symmetrically define $V_{MAJ5}$, assuming a pre-set of 1.

\noindent{\bf XOR:}
XOR is an especially useful gate for comparison, however, a
single-gate \arch\ implementation is not possible: In this
case we need $Out$ (not) to switch for 00 and 11, but not for 01 and
10, if the pre-set is 1 (0). However, due to $I_{00}>I_{01}=I_{10}>I_{11}$,
and assuming a pre-set of 1, we cannot let both  $I_{00}$ and $I_{11}$ remain
higher than $I_{crit}$ (such that $Out$ switches), while $I_{01}=I_{10}$ remain
lower than $I_{crit}$ (such that $Out$ does not switch). The same observation
holds for a pre-set of 0, as well.

\begin{table}[ht]
	\vshrink{0.3}
	\centering
		\begin{tabular}{c|c||c|c||c}
					&			& $S_1$=		  & $S_2$=		& $Out$= \\
			{$In_0$} & {$In_0$} & NOR($In_0$,$In_1$) & COPY($S_1$)& TH($In_0$,$In_1$,$S_1$,$S_2$) \\
			\hline
			0 & 0 & 1 & 1 & 0 \\
			\hline
			0 & 1 & 0 & 0 & 1 \\
			\hline
			1 &  0 & 0 & 0 & 1 \\
			\hline
			1 &  1 & 0 & 0 & 0
		\end{tabular}
	\caption{XOR implementation\label{tbl:exor}}
  \vshrink{0.7}
\end{table}

We can implement XOR using a combination of universal \arch\ gates such as NOR.
Thereby each XOR takes at least 4 steps (i.e., logic evaluations). 
For pattern matching, we will rely on a more efficient 3-step implementation (Table ~\ref{tbl:exor}): 
In {\em Step-1}, we compute $S_1$=NOR($In_0$,$In_1$). In {\em Step-2}, we perform $S_2$=COPY($S_1$). In
the final {\em Step-3}, we invoke 
a 4-input thresholding TH function, which
renders a 1 only if its inputs contain more than two zeros: $Out$ =
TH($In_0$,$In_1$,$S_1$,$S_2$).
TH has a pre-set of 0, and the operating principle is very similar to the
majority gates except that TH accepts an even number of inputs. We can further
optimize this implementation, and fuse {\em Step-1} and {\em Step-2} by
implementing NOR as a two-output gate.    

\begin{figure*}[!tp]
  \centering
  \includegraphics[width=\textwidth]{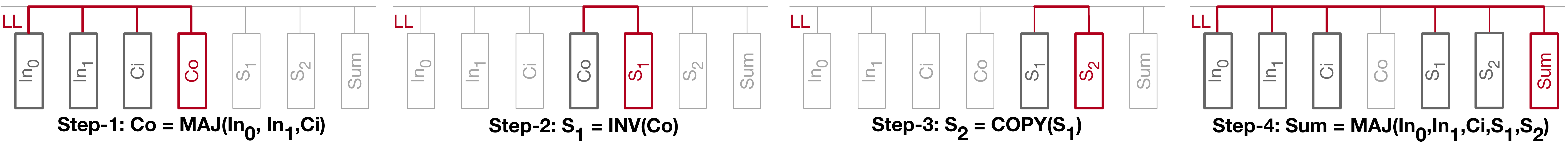}
 \vshrink{0.4}
  \caption{Full adder implementation~\cite{cram}. Output of each gate is depicted in red.\label{fig:FA}}
 \vshrink{0.5}
\end{figure*}

\noindent{\bf Full Adder:}
A full adder has three inputs: $In_0$, $In_1$, and carry-in $C_i$. The two
outputs are $Sum$ and the carry-out $C_o$. 
Like other logic functions, we can implement this adder using NOR gates.
However, an implementation based on a pair of MAJ gates
reduces the required number of steps significantly~\cite{Augustine11}.
Fig.\ref{fig:FA} provides a step-by-step overview:
\\ {\em Step-1}: $C_{o}$ = MAJ($In_0$,$In_1$,$C_i$)
\\ {\em Step-2}: $S_1$ = INV($C_{o}$)
\\ {\em Step-3}: $S_2$ = COPY($S_1$)
\\ {\em Step-4}: $Sum$ = MAJ($In_0$,$In_1$,$C_i$,$S_1$,$S_2$)


\subsection{Reconfigurability}
\label{sec:config}
\noindent Invoking a logic gate within the \arch\ array translates into pre-setting the
output, connecting all cells participating in computation to {\em LL} by setting
the corresponding {\em LBL}s (while keeping the {\em WL} at 0), grounding {\em
BSL} of the output, and setting {\em BSL}s of the inputs to  $V_{gate}$, which
depends on the type of the logic gate. Therefore, modulo output pre-set, the
complexity of reconfiguration is very similar to the complexity of addressing in
the memory array. 
%
\arch\ is reconfigurable along two dimensions:
\begin{list}{\labelitemi}{\leftmargin=1em}
\item Each cell can serve as an input or as an output for a logic gate depending on
  the computational demands of the workload within the course of execution.
\item For a fixed input-output assignment, the logic function itself is
  reprogrammable. For example, we can reconfigure the gate from
  Fig.\ref{fig:basicArray}(b)/(c) to implement another function than
  NOR by simply changing $V_{gate}$, to, e.g., $V_{NAND}$ (and applying a
  different output pre-set, as need be).
\end{list}

By default, \arch\ acts as an MRAM array.
A dedicated 
architecturally visible
set of registers keep the configuration bits to program
\arch\ cells as logic gate input/outputs. 
These configuration bits capture not only the physical location in the
array, but also whether the cell represents an input or an output, the pre-set
value for the output, and $V_{gate}$. 
A fixed or floating portion of
the \arch\ array can keep these configuration bits as part of the machine state,
as well. 

\subsection{Row-level Parallelism}
\label{sec:par}
\noindent 
\arch\ can perform only one type of logic function in a
row, at a time. This is because there is only one {\em LL} that spans the entire
row, and any cell within the row to participate in computation gets directly
connected to this {\em LL} (Section~\ref{sec:fuse}). 
On the other hand, the voltage levels on {\em BSL}s determine the type of the
logic function, where each {\em BSL} spans an entire column. 
Furthermore, in each row, each {\em LBL} -- which connects a cell participating in
computation to {\em LL} -- also spans an entire column.
Therefore, all rows can
perform the very same logic function in parallel, on the same set of columns.   

In other words, \arch\ supports a special form of {\em SIMD (single instruction
multiple data)} parallelism, where {\em instruction} translates into {\em logic
gate/operation}; and {\em data}, into {\em input cells in each row, across
all rows, which span the very same columns}.  

To summarize, \arch\ can only have either all rows computing in parallel, or the entire array serving as memory. Regular memory reads and writes cannot proceed simultaneously with computation.
Large scale pattern matching problems can greatly benefit from this execution model, as we are going to demonstrate in the following. 

\subsection{System Integration}
\label{sec:si}
\noindent 
\arch\ can serve as a stand-alone compute engine or a co-processor
attached to a host processor.  
Following the near-memory processing taxonomy from~\cite{nmpTax}, due to the
reconfigurability (Section~\ref{sec:config}), both \arch\ design points still fall
into the ``programmable'' class.
%
A classic system has to specify how to offload both {\em computation and data}
to the co-processor, and how to get the results back from the co-processor. For
a \arch\ co-processor, we do not need to communicate data values -- instead, the
\arch\ array requires (ranges of) data addresses to identify the data to
process, and the specification for computation, i.e., which function to perform
on the corresponding data.
In Section~\ref{sec:compiler} we will detail this interface.

\subsection{Spatio-Temporal Scheduling}
\label{sec:opt}
\noindent The goal of classic memory data layout optimizations is to perform as
many computations as possible per unit data delivered from the memory to the
processor, as the data communication between the processor and the memory
represents the bottleneck. 
\arch, on the other hand, brings compute capability to the data to be processed.
The goal becomes {\em minimizing the direct physical distance between the cells
participating in computation}.  Considering that an output cell can serve as an
input cell in subsequent steps of computation, the physical location of the
cells carrying the input data for subsequent steps can dynamically change as
computation proceeds.

This optimization problem gives rise to two strongly correlated sub-problems:
the layout of data to be processed in the memory array, and the spatio-temporal
scheduling of computations within the array.  In this regard, the optimization
problem has many analogies to floor-planning and placement algorithms deployed
in the computer aided design of digital systems, which aim to minimize the
``distance'' (in terms of wire length) between interconnected circuit blocks.
In \arch\ context, ``interconnected blocks'' translate into interconnected cells
(over {\em LL}) participating in computation (Section~\ref{sec:fuse}). We will
look closer into this effect in Section~\ref{sec:overhead}.

\arch\ hence features a unique trade-off between data replication
and
parallelism: Due to the internal array structure, (unless replicated), the
same cell can only participate in one computational step at a time, which may impair
opportunities for parallel execution.
Data replication can unlock more parallelism in such cases, at the expense of a
larger memory footprint.


\section{Spintronic Pattern Matching}
\label{sec:patt}  

\noindent 
Pattern matching is a key computational step in  large-scale data analytics. The most common form by far is character string matching, which involves repetitive search over very large databases residing in memory. Therefore, compute substrates such as \arch, that collocate logic and memory
to avoid the latency and energy overhead of expensive data transfers, have great
potential. Moreover, comparison operations dominate the
computation, which represent excellent acceleration targets for \arch. As a representative and important large-scale string matching problem, in the following, we will
use DNA sequence alignment~\cite{bigDataGenomics} as a running example, and
expand \arch's evaluation to other string matching benchmarks in Section~\ref{sec:eval}.

At each step, {DNA sequence alignment} tries to map a short character string to (the most similar substring of) an orders-of-magnitude-longer character string, and repeats this process for a very large number of short strings, where the longer string is fixed and acts as a reference.  For each string, the
characters come from the alphabet A(denine), C(ytosine), G(uanine), and T(hymine).  

The long string represents a complete genome; short strings, short DNA sequences (from the same species). The goal is to extract the region of the reference genome to which the short DNA sequences correspond to.  In the following, we will refer to each short DNA sequence as a {\em pattern}, and the longer
reference genome as {\em reference}.

Aligning each pattern to the most similar substring of the reference usually involves character by character comparisons to derive a {\em similarity score},
which captures the number of character matches between the pattern and the
(aligned substring of the) reference. 
Improving the throughput performance in terms of {\em number of patterns
processed per second} in an energy-efficient manner is especially challenging,
considering that a representative reference (i.e., the human genome) can be
around $10^9$ characters long, that at least 2 bits are necessary to encode each
character, and that a typical pattern dataset can have hundreds of millions
patterns to match~\cite{ajay}, where  
\arch\ can help due to reduced data transfer overhead and row-parallel
comparison/similarity score computations.   

Besides pattern matching, DNA sequence alignment algorithms include pre- and
post-processing steps, which typically span (input) data transformation for more
efficient processing, search space compaction, or (output) data re-formatting. 
In the following, we will only focus on the pattern matching operations, the
execution time share of which can easily reach 88\% in highly optimized GPU
implementations of popular alignment algorithms~\cite{klus2012barracuda}
\footnote{
For this implementation of the common BWA algorithm, the time share of the pattern matching kernel,
  inexact\_match\_caller, increases from 46\% to  88\%, as the number of base
  mismatches allowed (an input parameter to the algorithm) is varied from one to
  four (both representing typical values).  
}.

Mapping any computational task to the \arch\ array translates into
co-optimizing the data layout, data representation, and the spatio-temporal
schedule of logic operations, to make the best use of \arch's row-level
parallelism (Section~\ref{sec:par}).
This entails distribution of the data to be processed, i.e., the
reference and the patterns, in a way such that each row can perform independent
computations. 

The data representation itself, i.e., how we encode each character of the
pattern and the reference strings, has a big impact on both the storage and the
computational complexity. Specifically, data representation dictates not only
the type, but also the spatio-temporal schedule of (bit-wise) logic operations. 

Spatio-temporal scheduling should also take intermediate results during
computation into account, which may or may not be discarded (i.e., overwritten),
and which may or may not overwrite existing data, as a function of the algorithm
or array size limitations.

\subsection{Data Layout \& Data Representation}
\label{sec:data}
\begin{figure}[h]	
	\vshrink{0.2}	
	\centering
	\includegraphics[width=0.43\textwidth]{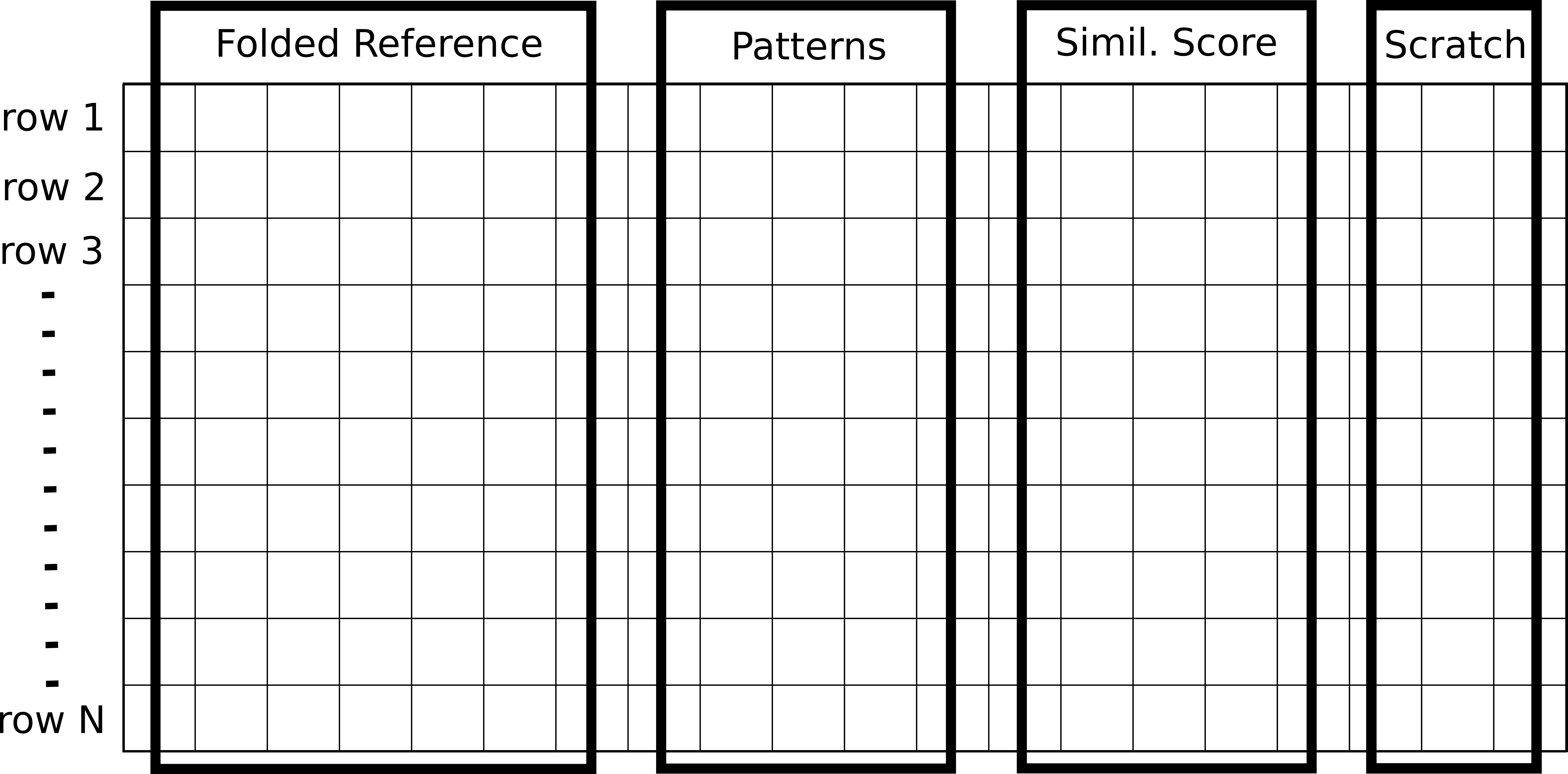}
	\vshrink{0.1}
	\caption{Data layout per \arch\ array.}
	\label{fig:layout}
	\vshrink{0.3}
\end{figure}

\noindent 
Without loss of generality,
we use the data layout
captured by Fig.~\ref{fig:layout}, by folding the long reference over multiple
\arch\ rows. Each row has four dedicated compartments to accommodate a fragment
of the folded reference; one pattern; the similarity score (for the pattern when
aligned to the corresponding fragment of the reference); and intermediate data
(which we will refer to as {\em scratch}). The same format applies to each row,
for efficient row-parallel processing. Each row contains a different fragment
of the reference.  

We determine the number of columns allocated for each of the four compartments,
as follows:
In the DNA alignment problem, the reference corresponds to a genome, 
therefore, can be very long. The species determine the length. As a case
study for large-scale pattern matching, in this paper we will use
approx. 3$\times10^9$ character-long human genomes. 
Each pattern, on the other hand, represents the output from a DNA sequencing
platform, which biochemically extracts the location of the
four characters (i.e., bases) in a given (short) DNA strand. Hence, the sequencing
technology determines the maximum length per pattern, and around 100
characters is typical for modern platforms processing short DNA strands~\cite{illumina}.
The size of the similarity score compartment, to keep the
character-by-character comparison results, is 
a function of  
the pattern length. 
Finally, the size of the scratch compartment depends on both the reference
fragment and pattern length. 

While the reference length and the pattern length are problem specific
constants, the (reference) fragment length (as determined by the folding
factor), is a \arch\ design parameter. By construction, each fragment should be
at least as long as each pattern. The maximum fragment length, on the other
hand, is limited by the maximum possible \arch\ row length, considering
the maximum affordable capacitive load (hence, RC delay) on row-wide control
lines such as {\em WL} and {\em LL}. However, row-level parallelism favors
shorter fragments (for the same reference length). The shorter the fragments,
the more rows would the reference occupy, and the more rows, hence regions of
the reference, would be ``pattern-matched'' simultaneously.

\ignore{
The length of the arrows in the figure is not representative of the number of
cells in a particular row they require. Organizing data in this fashion allows
us to execute the same sequence of operations on the data in each row at the
same time, thereby boosting the throughput of the array. The critical component
in this layout is the length of the reference segments and the scores. The
longer the reference segment stored in each row, the less number of arrays would
be required (assuming all arrays ahre the same column height). However, making
the reference segments shorter and increasing the number of rows in one array
i.e. increased column height would increase the number of queries processed at
the same time since all rows have queries which they align with the reference
segment at that corresponding row. The query length determines how many bits
i.e. CRAM cells are required to be used in logic execution and to store the
intermediate results. Furthermore, the length of the query sequence determines
what is the length of scores at each base location of the reference segment,
which is equal to $log_2(query\_length)$ rounded to the next integer. The amount
of bits required to store the scores actually depends on how the underlying
algorithm works. For instance, if rows store all alignment scores generated by
shifting the query sequence over reference segment (one base character at a
time), the total number of scores will be equal to the difference between the
reference segment length and query sequence. The number is usually significantly
high and contributes the most to the required number of cells in a row.  An
alternative approach would be use a score buffer structure in the accelerator
pipeline which reads scores from the rows when the computations are done. This
would introduce an idle window of time when the arrays will wait for the buffer
to read the scores from the rows, one at a time. Since the read time of one row
is ~0.5 ns, the total idle time is just this time multiplied by the array
height. This approach has the advantage of having a much reduced number of cells
per row to store the alignment score (one score per row) at the expense of
buffer read time overhead.       
}

For data representation, we simply use 2-bits to encode the four
(base) characters, hence,
each character-level comparison entails two bit-level
comparisons.  

\ignore{
The first issue to consider when mapping reference
sequence and pattern to CRAM is how to represent the nucleotide bases in terms
basic building blocks of CRAM-CRAM cells. This directly affects how much memory
would be required to store a given reference sequence and pattern in the array.
Every DNA sequence read by NGS platform is a sequence of finite length which is
a combination of the four base elements- A, T, C and G. How to represent these
characters in the CRAM structure is dependent on the range of operations to be
performed on these and how those operations are carried out in CRAM context. For
example, if two bits (CRAM cells) are required to be EXOR-ed to each other, the
detail of the gate i.e. sequence and number of gates connected to each other to
execute EXOR function, dictates what is the preferred encoding scheme. To
represent 4 base characters, we use 2 bits per character, for instance- 00 (A),
01 (T), 10 (C) and 11 (G). Each bit is stored in one CRAM cell. Therefore, when
matching two base characters (one from reference sequence and another form
pattern) the matching is done bit by bit i.e. cell by cell. Using only 2 bits
per base character makes it possible to store reference and base characters
using minimum number of CRAM cells. 
}

\subsection{Proof-Of-Concept \arch\ Design}
\label{sec:design}
\arch\ comprises two computational phases, which Algorithm \ref{algo1} captures at the row-level: {\em match}, i.e., {\em aligned bit-wise comparison} and {\em similarity score computation}. As each row
performs the very same computation in parallel, in the following, we will detail
row-level operations. 

\begin{algorithm} 
 \caption{2-phase pattern matching at row-level} \label{algo1}
  \small
  \begin{algorithmic} 
   \STATE $loc= 0$ \WHILE{$loc< len$($fragment$)$ - len$($pattern$)}
  \STATE {\bf Phase-1: Match (Aligned Comparison)} \\  
  align pattern to location $loc$ of reference fragment;\\ 
  (bit-wise) compare aligned pattern to fragment \\  
  \STATE {\bf Phase-2: Similarity Score Computation} \\  
  count the number of character-wise matches;\\
  derive similarity score from count
  \STATE $loc++$		
  \ENDWHILE \end{algorithmic}
\end{algorithm}

In Algorithm~\ref{algo1}, $len$($fragment$) and $len$($pattern$) represent the (character) length of the reference fragment and the pattern, respectively; and $loc$, the index of the fragment string where we align the pattern for comparison.
%
The computation in each row starts with aligning the fragment and the pattern string, from the first
character location of the fragment onward. For each alignment, a bit-wise comparison of the
fragment and pattern characters comes next. The outcome is a $len$($pattern$) bits long
string, where a 1 (0) indicates a character-wise (mis)match. We will refer to
this string as the {\em match string}. Hence,   the number of 1s in the match string acts as a measure for how similar the fragment and the pattern are, when aligned at that particular character location ($loc$ per Algorithm~\ref{algo1}).

A reduction tree of 1-bit adders counts the number of 1s in the match string to derive the similarity score. Once the similarity score is ready, next iteration starts. This process continues until the last character of the pattern reaches the last character of the fragment, when aligned.  

\begin{figure}[h]	
	\vshrink{0.6}
	\subfloat[]{
		\centering
		\includegraphics[width=0.21\textwidth]{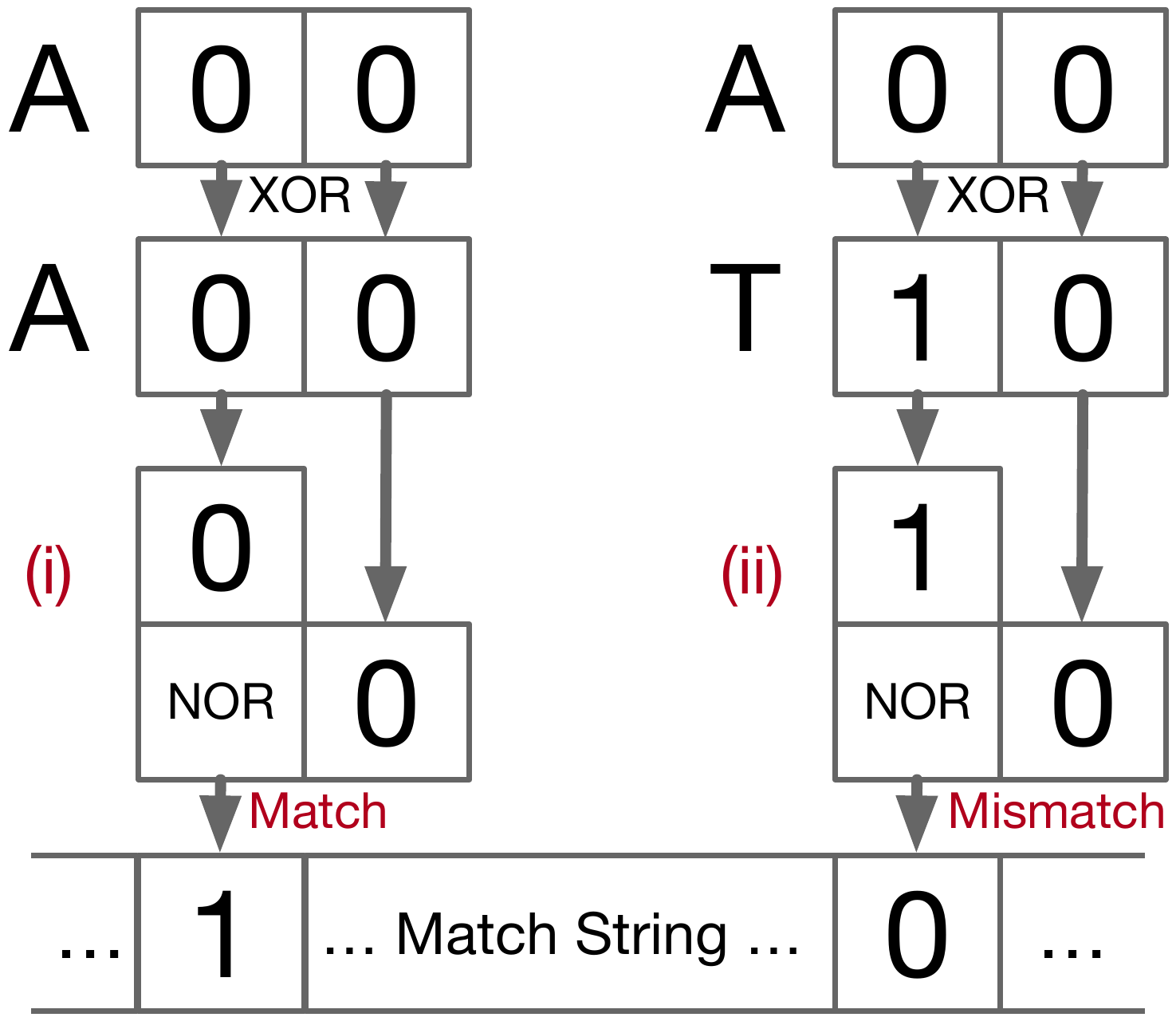}
		\label{fig:data_op_match}
	}
	\hfill
	\subfloat[]{
		\centering
		\includegraphics[width=0.21\textwidth]{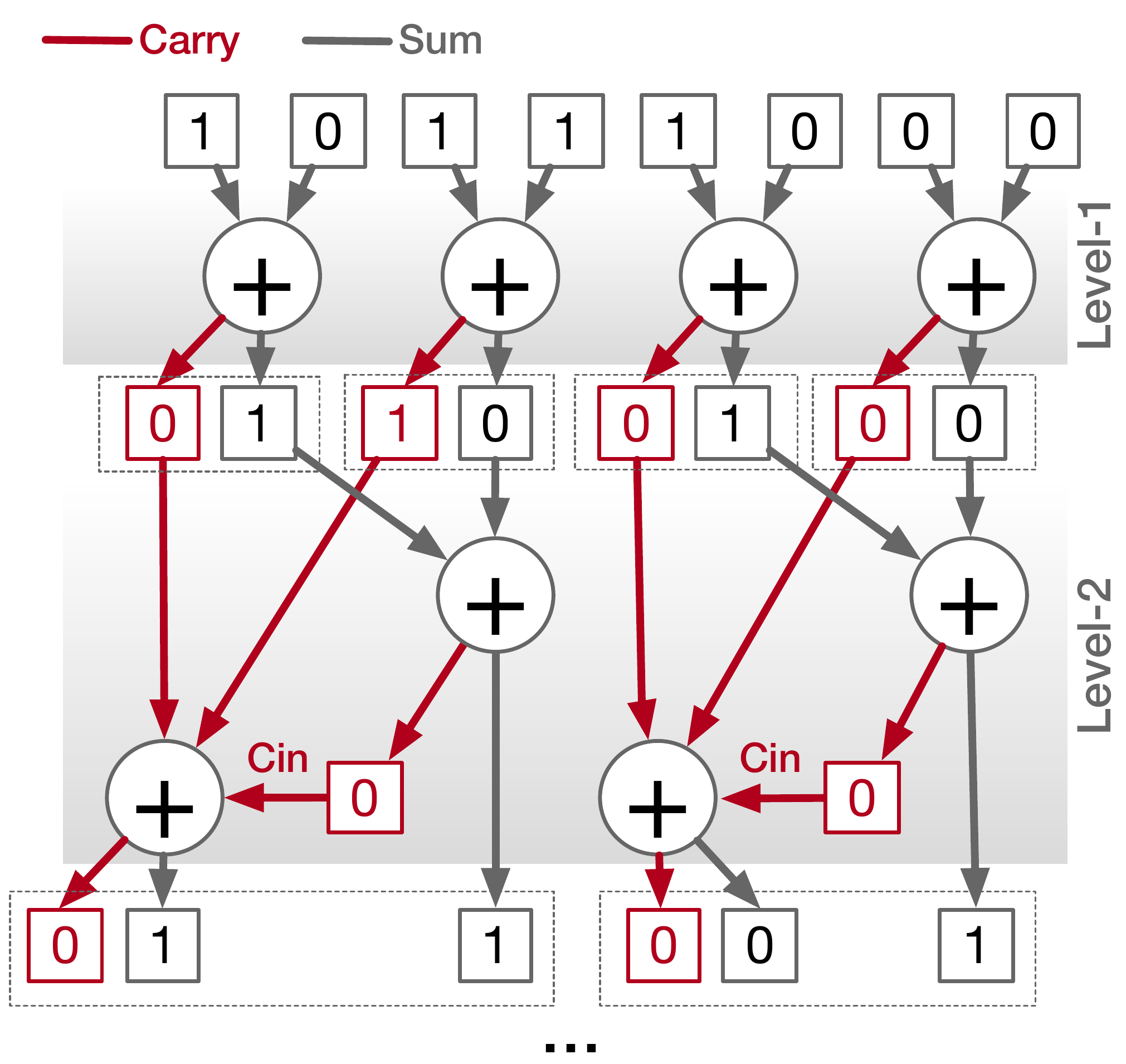}
		\label{fig:data_op_score_comp}
	}
	\vshrink{.3}
	\caption{Aligned bit-wise comparison (a), and adder
		reduction tree used for similarity score computation (b).}
	\label{fig_data_op_reduction_adder}
	\vshrink{0.3}	
\end{figure}

\noindent{\bf Phase-1 (Match, i.e., Aligned Comparison):} 
Each aligned character-wise comparison gives
rise to two bit-wise comparisons, each performed by an 2-input XOR gate. 
Fig.\ref{fig:data_op_match} provides an example, where we compare the base
character `A' (encoded by `00') of the fragment with the base character `A' (i),
and `T' (encoded by `10') (ii), of
the pattern. A 2-input NOR gate converts the 2-bit comparison outcome to a
single bit, which renders a 1 (0) for a character-wise (mis)match. Recall that a
NOR gate outputs a 1 only if both of its inputs are 0, and that an XOR gate
generates a 0 only if both of its inputs are equal.
The implementation of these gates follows from Section \ref{sec:bb}.

\arch\ can only have one gate active per row at a time (Section \ref{sec:par}).
Therefore, for each alignment (i.e., for each $loc$ or iteration of Algorithm \ref{algo1}),
such a 2-bit comparison takes place $len$(${pattern}$) times in each row, one after
another. Thereby we compare all characters of the aligned pattern to all characters of the 
fragment, before moving to the next alignment (at the next location $loc$ per
Algorithm \ref{algo1}).
That said, each such 2-bit comparison takes place in parallel over all rows,
where the very same columns participate in computation. 

\vshrink{-0.1}
\noindent{\bf Phase-2 (Similarity Score Computation):}
For each alignment (i.e., iteration of Algorithm \ref{algo1}), once all bits of the
match string are ready -- i.e., the character-wise comparison of the
fragment and the aligned pattern string is complete for all characters, we count the
number of 1s in the match string to calculate the similarity score.
A reduction tree of 1-bit adders performs the counting,
as captured by Fig.\ref{fig:data_op_score_comp}, with the carry and sum
paths shown explicitly for the first two levels.
The top row corresponds to the contents of the match string; and each 
$\oplus$, 
to a 1-bit adder from Section \ref{sec:bb}. 


$len$(${pattern}$), the pattern length in
characters, is equal to the match string length in bits.  
Hence, the number of bits required to hold the final bit-count (i.e., the
similarity score)
is 
$N=\lfloor{log_2 len(pattern)}\rfloor + 1$.
A naive implementation for the addition of $len$(${pattern}$) number of bits requires 
$len$(${pattern}$) steps, with each step using an
$N$-bit adder, to generate an 
$N$-bit partial sum towards the
$N$-bit end result.
For a typical pattern length of around 100~\cite{illumina}, this translates into 
approx. 100 steps, with each step performing a
$N=7$ bit addition.
Instead, to reduce both the number of steps and the operand width per step, we
adopt the reduction tree of 1-bit adders from Fig.\ref{fig:data_op_score_comp}.
Each level adds bits in groups of two, using 1-bit adders.  
\ignore{
There is no carry at
the first level, hence, the number of 1-bit 
adders becomes 
half the number of bits in the match string.  The second level requires
$\lceil{l(pattern)/4}\rceil$ 2-bit adders (considering the carry out generated at
the first level), and so on and so forth.
%
}
For a typical pattern length of around 100~\cite{illumina}, we thereby reduce
the complexity to 188 1-bit additions in total.

\vshrink{-0.1}
\noindent{\bf Data Output:}
Each iteration of Algorithm \ref{algo1} at the end of Phase-2 generates a new
similarity score in each row.
One approach is, in each row, keeping the
similarity score for all iterations. This requires
$O$($len$($fragment$) $\times N$) bits per row, as 
each score takes
$N$ bits, and one pass of Algorithm~\ref{algo1} takes $O$($len$($fragment$)) iterations.
An alternative approach, 
to trade storage complexity for execution time, 
is using a dedicated score buffer 
at the array periphery (similar to the row buffer in main memory)
to have each new score (per row) read out at the end of Phase-2, before the next
iteration starts. In this case,
each row only has space for one similarity score (of $N$ bits).
This introduces an idle time window before the next iteration can fire, since  
we can only read out one score (from each row) at a time. Still, 
considering the overhead of pre-setting output cells to prepare for the next
iteration, we can mask the overhead of read-outs. This trade-off strongly
depends on the values of the fragment and pattern lengths.    

In either case, \arch\ annotates each score with the row number and column
number (in the folded reference) where the respective pattern was aligned. 
The column number simply corresponds to $loc$ from Algorithm \ref{algo1}.
The host processor can use this information to extract the maximum-score alignment, or
to rank alignments for further analysis.

\vshrink{-0.1}
\noindent{\bf Assignment of Patterns to Rows:}
In each \arch\ array we can process a given pattern dataset in different ways.
We can assign a different pattern to each row, where a different fragment of the
reference resides, or distribute the very same pattern across all rows. Either
option works as long as we do not miss the comparison of a given pattern to all
fragments of the reference.  In the following, we will stick to the second
option, without loss of generality. This option eases capturing alignments
scattered across rows (i.e., where two consecutive rows partially carry the most
similar region of the reference to the given pattern).  A large reference
can also occupy multiple arrays and give rise to scattered alignments at array
boundaries, which row replication at array boundaries can address.


\ignore{
Although the proposed design and the evaluation
presented are intended for aligning the patterns with reference sequence
considering exact match, this may not be the case in practice. In real systems,
the matches might not always align with the reference segment and therefore
require fragmented matches of patterns, which extend from one reference segment
to the next. In absence of such scenario, the reference segments are
non-overlapping which suffices the similarity search of a total of
$(ref\_seg\_length - pattern\_length)$ matching along the reference segment
length. However, in case of a partial match we would need to go beyond that
segment length, toward next segment, to find the similarity score. The
straightforward solution to this problem is to use overlapped reference segments
in successive row of SpinCM array. Figure~\ref{fig_frag_match} illustrates this
scenario. Steps $i$, $i+1$ and $i+2$ are performing fragmented similarity search
with two reference segments which are stored in the subsequent row of SpinCM
array.  

\begin{figure}[!]	\centering
\includegraphics[width=0.47\textwidth]{Figs/fragmented_match_March5.png}
\caption{Fragmented matching of pattern.} \label{fig_frag_match} \end{figure}

This enables the similarity search to virtually extend beyond the any
participating row. The downside of this approach is, of course, the redundant
computations in each row. Since, in current form of SpinCM, inter-row
computation and communication are not considered, we can achieve fragmented
similarity search in this approach. The amount of extra base characters in each
row is equal to $pattern\_length -1$. 
}

\ignore{

\subsubsection{Latency Model} \label{anamodel_latency} Figure
~\ref{fig_latency_model} shows the timing diagram corresponding to the compute
phases from Algorithm ~\ref{algo1}. The write bits refers to the writing of
pattern bits on each row of each array. The latency associated with bitline
activation is considered here as well. Once the bits are written to an entire
array or system of arrays, the \textit{match phase} begins. The operations
begins with matching the heads of pattern and the corresponding reference
segments. The \textit{match phase} is an iterative process which loops over the
entire length of the reference segment on a particular rows. Before performing
the match operations, the corresponding output bits are preset to the required
state (0 or 1) and the bitlines associated with the input and output cells are
activated. Upon exiting this loop, it has a string of 0's and 1's in the output
cells. The bits are counted with the help of a series of adders in reduction
tree like approach. In each of the adder steps in \textit{compute score phase},
the corresponding output cells are preset and the bitlines are activated. When
the \textit{score compute phase} is complete, there is a window of wait time
introduced to allow for the scores to be read from the rows of an array (buffer
read). The whole process iterates over the entire length of the reference
segments, with the head of pattern aligned with the next base character-
starting from the head of the corresponding reference segment. Once the pattern
reaches the end location on reference segment i.e. $reference\_segment\_length -
pattern\_length$, the similarity search process is complete. The latency (and
throughput) of an array is simple the latency of any row since all rows perform
similarity search in parallel.                  

\begin{figure*}[!]	\centering
\includegraphics[width=0.95\textwidth]{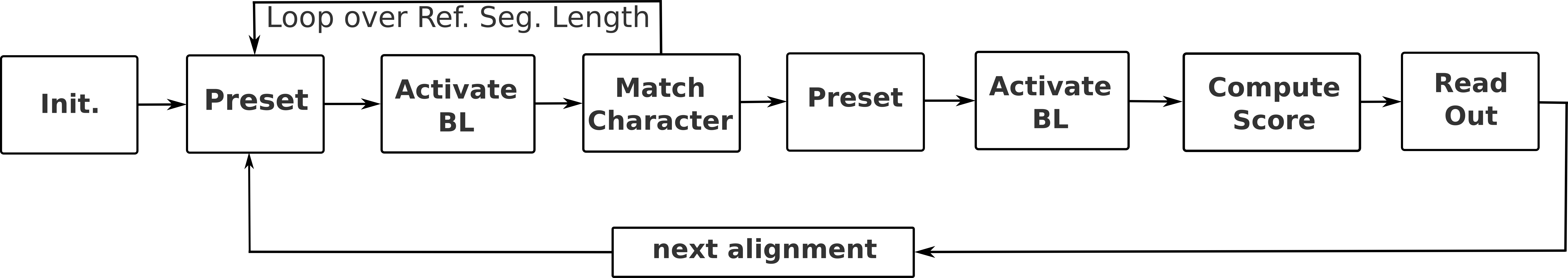} \caption{Process
flow diagram} \label{fig_latency_model} \end{figure*}

\subsubsection{Energy Model} \label{anamodel_energy} The energy model is derived
from the latency model, with the exception that the energy of an entire array
has contributions from individual rows in that array. The steps/operations shown
in Figure~\ref{fig_latency_model} is repeated for every row of an array, and the
aggregated energy contributes toward the total the energy of the design. This
energy multiplied by total number of arrays required to hold the reference
sequence is the total energy consumed by the designed system.      

\subsection{Data Operations} The operations required to be performed on the data
in each row is dependent on the Pseudocode shown in Algorithm \ref{algo1}. In
this section, we present the data operations required to generate the similarity
scores for patterns. The two computational phases- \textit{match phase} and
\textit{score compute phase} require basic computational functions such as EXOR,
bit-wise addition etc. These computational functions are further comprised of
fundamental logic functions e.g. NOR, NOT etc. 

\subsubsection{Match Phase} At the first step of alignment, bit-wise match
operation is performed on each of the two bits that represent a base character.
This is done with an equivalent exclusive-OR (EXOR) operation. Since each base
character is represented with multiple bits and we need to see a representative
1 (or 0, depending on the design), the multiple outputs from EXOR operations
have to be merged into a single output. This step is done using a NOR operation.
In case of a match, corresponding encoded bits (of base characters) would
generate 0s which will generate, with NOR function, 1 as the match output.
Otherwise, the final outcome will be 0, to represent the presence of a mismatch.
Figure~\ref{fig_data_op_match} shows an example case where base character 'A' is
compared with characters 'A' and 'T', illustrating a match and a mismatch
respectively. The encoding used here is for this example case only and does not
present any unique solution.

\begin{figure}[!]	\centering
\includegraphics[width=0.45\textwidth]{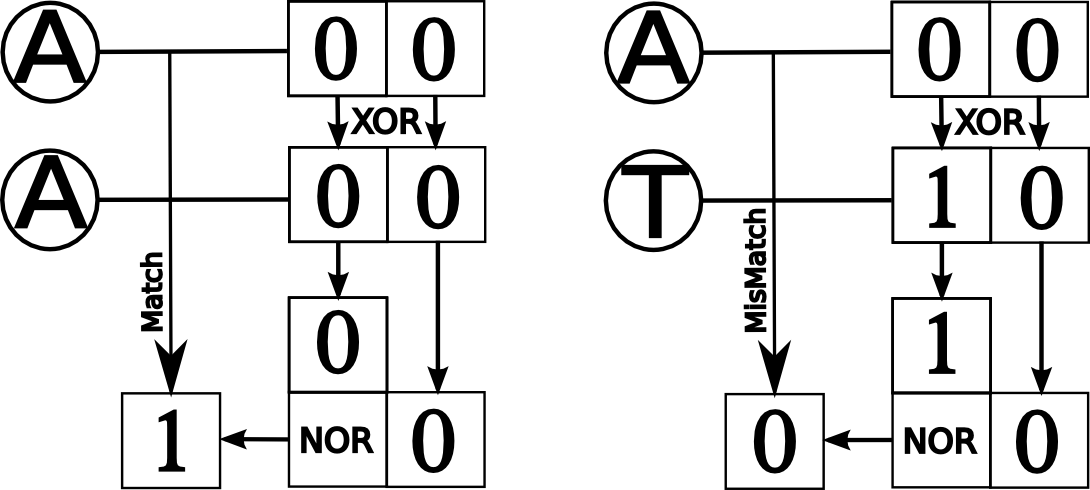}
\caption{Data operations during \textit{match phase}} \label{fig_data_op_match}
\end{figure}

\subsubsection{Score Compute Phase} Once all entires in the match string is
computed i.e. the difference in base characters are resolved for one iteration,
the adder operations are performed on that string. The purpose of these addition
operations are to count the numbers of 1's (or 0's) to accumulate the similarity
score at a particular location along reference sequence segment. The design of
the adder section is straightforward since it follows 1-bit full adder
technique, however, with a custom combination of gates. This involves using
MAJ(ority)3 , MAJ(ority)5 and buffer gates, as well as inverters (see
section~\ref*{section_adder} for more details). The key concept here is to
understand how to find the similarity score from a match string. The number of
bits required to hold the addition result is $log_2(pattern\_char\_count)$
(rounded to the next integer). To perform an addition of match\_string length of
bits, the number of additions would be equal to the $match\_string\_length - 1$
where each addition will require $log_2(pattern\_char\_count)$ number of adders
with padded bits. 
$log_2(pattern\_char\_count)$ number of bits to get the final similarity score.
For a pattern length of 100, the required number of 1-bit adders is 700. This is
a tremendous amount of steps to be performed which would hurt the performance
(both latency and energy) of the similarity search algorithm. We optimized this
approach to reduce the number of addition stages and the width of each addition. 

\begin{figure}[!]	\centering
\includegraphics[width=0.35\textwidth]{Figs/adder_tree_march1.png} \caption{Data
operations during \textit{score compute phase}} \label{fig_data_op_score_comp}
\end{figure}

The score computation is performed using a reduction adder tree like structure
consists of 1-bit full adders (FA). At the first stage, the number of 1-bit
adders is half (rounded to the next integer) of the number of entries in the
match string. The subsequent round will require $1/4$ of the
match\_string\_length number of 2-bit adders, and so on.
Figure~\ref{fig_data_op_score_comp} shows an example of this optimization. The
match string contains 4 matches i.e. 1s and the reduction tree adds the bits in
groups of two using 1-bit adders. The resultant carry and sum bits are again
added using two 2-bit adders i.e. four 1-bit adders to produce the binary sum
$0100$. The $C$ shown here is the carry from the addition operation which is
utilized by the next stage of additions. The total number of adders (and the
bit-width of adders as well), using such a tree structure, is reduced
significantly. For instance, for a pattern length of 100 base characters, the
match string has 100 entries. Straightforward or naive addition will require 700
7-bit addition operations to be performed. However, with our optimization, the
required number of adders is 188 where each adder is an 1-bit adder. This
significantly reduces the latency and energy of \textit{score compute phase}.  

Although the proposed design and the evaluation
presented are intended for aligning the patterns with reference sequence
considering exact match, this may not be the case in practice. In real systems,
the matches might not always align with the reference segment and therefore
require fragmented matches of patterns, which extend from one reference segment
to the next. In absence of such scenario, the reference segments are
non-overlapping which suffices the similarity search of a total of
$(ref\_seg\_length - pattern\_length)$ matching along the reference segment
length. However, in case of a partial match we would need to go beyond that
segment length, toward next segment, to find the similarity score. The
straightforward solution to this problem is to use overlapped reference segments
in successive row of SpinCM array. Figure~\ref{fig_frag_match} illustrates this
scenario. Steps $i$, $i+1$ and $i+2$ are performing fragmented similarity search
with two reference segments which are stored in the subsequent row of SpinCM
array.  

\begin{figure}[!]	\centering
\includegraphics[width=0.47\textwidth]{Figs/fragmented_match_March5.png}
\caption{Fragmented matching of pattern.} \label{fig_frag_match} \end{figure}

This enables the similarity search to virtually extend beyond the any
participating row. The downside of this approach is, of course, the redundant
computations in each row. Since, in current form of SpinCM, inter-row
computation and communication are not considered, we can achieve fragmented
similarity search in this approach. The amount of extra base characters in each
row is equal to $pattern\_length -1$. 

}




\subsection{System Interface}
\label{sec:compiler}

\noindent We will next cover the \arch\ system stack to support in-memory
execution semantics for pattern matching.

\vshrink{-.1}
\noindent {\bf \arch\ Instructions:} In addition to conventional memory read and
write, \arch\ instructions cover computational building blocks for in-memory
pattern matching.  \arch\ instructions hence form two classes: data transfer
(read, write) and
computational (arithmetic/logic).  By construction, computational \arch\ instructions are {\em
block} instructions: two dimensional vector instructions, which operate {\em on
all rows} and {\em on a subset of columns} of an \arch\ array at a time.  Hence,
key operands for any computational \arch\ instruction are the column numbers of
the source(s) (i.e., input(s) to computation) and destination(s) (i.e.,
output(s) to computation). 
Depending on the size of the pattern matching problem, multiple \arch\ arrays
may be deployed in parallel. Therefore, the computational subset of \arch\
instructions facilitates gang-execution on all \arch\ arrays, as well. In the
following, we will generically use the term \arch\ {\em substrate} to refer to
all arrays participating in computation.
We also make the distinction between {\em macro}- and {\em micro}-instructions.
The set of micro-instructions covers actual bit-level operations performed in
the \arch\ substrate, while the set of macro-instructions forms the high-level
programming interface.

\vshrink{-.1}
\noindent {\bf Programming Interface:}
To match \arch's row-level parallelism, memory allocation and declaration of
variables (which represent inputs and outputs to
computation) happen at row granularity. Depending on the problem, a variable may
cover the entire row or only a portion. The following code snippet provides an
example, where an integer variable \texttt{x} gets written (assigned) to row
\texttt{r} and
column \texttt{c} in a \arch\ array (line 5):

\vshrink{-0.1}
%
{\footnotesize{ 
  1\indent$\mathtt{int~x~=~...;}$ \\
  \indent2\indent $\mathtt{...}$\\
  \indent3\indent $\mathtt{int~y;}$ \\
  \indent4\indent $\mathtt{preset}$($\mathtt{c,~ncell,~val}$); \\
  \indent5\indent $\mathtt{int_{pm}~x_{pm}~=~write_{pm}}$($\mathtt{x,~r,~c,~sizeof}$($\mathtt{x}$)); \\
  \indent6\indent $\mathtt{y~=~readdir_{pm}}$($\mathtt{x_{pm}}$); \\ 
}}

\vshrink{0.2}
In this case, besides \texttt{x} and \texttt{y}, \texttt{ncell}, \texttt{val},
\texttt{c} and \texttt{r} represent (already defined) integer values.  The
\arch-specific (composite) data type $\mathtt{int_{pm}}$ captures row and column
coordinates for each variable stored in the array. $\mathtt{x_{pm}}$ in line 5
keeps this information for variable \texttt{x}, after it gets written to row
\texttt{r}, from column \texttt{c} onwards, by the $\mathtt{write_{pm}}$
function.  The subsequent read in line 6, conducted by the
$\mathtt{readdir_{pm}}$, directly assigns the value of \texttt{x} to
\texttt{y}. \arch\ also features a read function, $\mathtt{read_{pm}}$,
which has a similar interface to $\mathtt{write_{pm}}$ with explicit row and
column specification. We consider each such function as a macro-instruction.

The \texttt{preset} function in line 4 presets \texttt{ncell} number of
(consecutive) cells, starting from column \texttt{c}, each to value \texttt{val}.
\arch\ features different variants of this function, including one to
gang-preset the entire scratch
area (Fig.~\ref{fig:layout}), and another where \texttt{val} is interpreted as a bitmask (of
\texttt{ncell} bits) rather than a single-bit preset value which applies over
the entire range of the specification.

Each pattern matching problem to be mapped to \arch\ features three basic
stages: 
\begin{itemize}
\item[(i)] Allocating and initializing the reference, pattern, and scratch
regions in each array (Fig.~\ref{fig:layout}); 
\item[(ii)] Computation;  
\item [(iii)] Collecting the pattern matching outcome. 
\end{itemize}

Variants of \texttt{preset} and
$\mathtt{write_{pm}}$ functions cover stage (i);  and variants of
$\mathtt{read(dir)_{pm}}$, stage (iii). Stage (ii) can take different forms
depending on the encoding of pattern and reference characters, but generally
primitives such as $\mathtt{add_{pm}}$($\mathtt{int~start, int~end, int_{pm}~result}$)
apply, which sums all cell contents between columns \texttt{start} and
\texttt{end}, on a per row basis, and writes the result back where
\texttt{result} points. $\mathtt{add_{pm}}$ macro-instruction can directly
implement Phase-2 from Algorithm\ref{algo1} to calculate the bit-count on the
match string (Section~\ref{sec:design}).

\ignore{
The compiler interface, shown in step-1,
	handles two basic tasks: data read/write and scratch location allocation in
	\arch, and operations on stored data. The scratch location is specified for
	one row only, and the same allocation is replicated across all rows. Scratch
	area is preset before computation which is basically write operation and to
	get the best performance, preset operations are optimized during compile
	time. 
  
The data allocation is provided by simple declaration of variable, at a row
granularity, since each row might contain different data. Within each row, the
data is placed at any location according to the application requirement. The API
takes in the variable locations in a row, to perform write operations. The
following code snippet copies i.e. writes variable x to specific \arch  row and
column. 
\begin{figure}[!htp]
	\centering
	\includegraphics[width=.47\textwidth]{Figs/code_snippet_01.pdf}
	\caption{Code snippet-01. 
		\label{fig:code_snippet_01}}
	\vshrink{0.3}
\end{figure}

Preset values depend on the type of operation to be performed. Therefore, preset
values As  code generation is in charge of composition of operations during is
in charge of the preset values are determined during compilation by the
compiler.  

These
micro-instructions are an extended set of low level instructions which are
received, decoded and executed by the \arch\ memory controller (SMC). Step-2
shows the micro-instructions for \textit{NAND}, \textit{NOR} and \textit{MAJ3}
operations. These micro-instructions only contain the type of operation and the
columns to connect as inputs and outputs.  For example, \textit{nand c3, c1, c2}
specifies column \textit{c3} as the output and column \textit{c1} and
\textit{c2} as inputs.       
Compiler
provides necessary API and data structure features to the developer- in order to
map the workload data to \arch\  and schedule \textit{in-memory} operations on
those data. 
}

\vshrink{-.1}
\noindent {\bf Code Generation:}
Code generation simply entails translating a sequence of 
macro-instructions to a sequence of micro-instructions
for the \arch\ memory controller (SMC) to drive the in-place computation.
Micro-instructions specify the type of operation and the
columns to connect as inputs and outputs.  For example, $\mathtt{nand}$($\mathtt{c_i, c_j,
c_k}$)
specifies column $\mathtt{c_i}$ as the output and column $\mathtt{c_j}$ and
$\mathtt{c_k}$ as inputs to form a NAND gate in the \arch\ array.  
The macro-instruction $\mathtt{nand_{pm}}$, on the other hand, performs the very
same operation on multi-bit operands (of width \texttt{ncell}):
$\mathtt{nand_{pm}}$($\mathtt{c_i, c_j, c_k, ncell}$).
In this case, $\mathtt{c_i}$, $\mathtt{c_j}$, and $\mathtt{c_k}$ still demarcate the starting
columns for the source and destination (\texttt{ncell} bit) operands.
$\mathtt{nand_{pm}}$ hence translates into
a sequence of \texttt{ncell} number of \texttt{nand} micro-instructions.
For $\mathtt{add_{pm}}$ type of
macro-instructions, on the other hand, a spatio-temporal
scheduling pass (Section~\ref{sec:opt}) determines the corresponding composition of micro-instructions. 
The goal is to maximize the throughput performance for the given data layout. This
usually translates into masking the overhead of presets or other types of writes
(per row) by coalescing when possible.
By construction, variants of \texttt{preset} 
macro-instruction trigger a
sequence of memory writes (as many as the number of rows), as at most one row can be written at a time.

\vshrink{-.1}
\noindent {\bf \arch\ Memory Controller (SMC):}
SMC orchestrates computation in the \arch\ substrate, and 
the communication with the host processor. 
\arch\ features an internal clock. During computation, SMC allocates each
micro-instruction a specific number of cycles to finish depending on the
operation and operand widths. This time window includes peripheral overheads and the
scheduling overhead due to SMC, besides computation. 
After the allocated time 
elapses (and unless an exception is the case), SMC fetches the next set of micro-instructions.
SMC features an instruction cache where micro-instructions reside until they are
issued to the \arch\ substrate. Before issue, 
SMC decodes the micro-instructions using a look-up table to initiate preset, and
subsequently, to 
set the appropriate voltage level on input BSL (as a function of the operation,
as explained in Section~\ref{sec:bb}), before 
activating the corresponding columns in the specified arrays for computation.              
The look-up table keeps the voltage level and the preset value for each bit-level
operation from Section~\ref{sec:bb}, which form a \arch\ micro-instruction.
No look-up table access is necessary for read and write operations.


\ignore{
- Code 
- Translation: use ISA extensions
  read 
  write
  op(...)
- Translation
  - Spatio-temporal optimization
- Runtime
  - Ctrller
  - SpinCM clock: start and end of comp

- Scratch allocation
- Preset schedule

- Disc
  - Programmability/reconfig.
  - Spatio-temporal optimization

  \begin{lstlisting}
  \scriptsize
  Hello world
  \end{lstlisting}

***************************************************
***************************************************

Figure~\ref{fig:systeminteg} {shows an overview of the system integration for
\arch. Being a processing-in-memory substrate, \arch\  enables computation within
memory as well, which is why the compiler interface has to support \textit{PIM}
operations in addition to conventional read and write operations. Compiler
provides necessary API and data structure features to the developer- in order to
map the workload data to \arch\  and schedule \textit{in-memory} operations on
those data. 

Operating system handles the API calls generated from the
executables and has \arch\  aware memory management to make the required data
layout to take place. The software stack also provides driver support which
includes \arch\  operation scheduling and sending the corresponding extended ISA
converted signals to the substrate controller. The controller interprets the
instruction stream and performs either memory read/write operations or data
processing on the stored data.}          

\begin{figure}[!htp]
	\centering
	\includegraphics[width=.47\textwidth]{Figs/compiler_overview.pdf}
	\caption{SpinPM System Integration. 
	\label{fig:systeminteg}}
	\vshrink{0.3}
\end{figure}

\textbf{Compiler Interface:} {The compiler interface, shown in step-1, handles
	two basic tasks: data read/write and scratch location allocation in \arch\,
	and operations on stored data. The scratch location is specified for one row
	only, and the same allocation is replicated across all rows. Scratch area is
	preset before computation which is basically write operation and to get the
	best performance, preset operations are optimized during compile time. 
	
The data allocation is provided by simple declaration of variable, at a row
granularity, since each row might contain different data. Within each row, the
data is placed at any location according to the application requirement. The API
takes in the variable locations in a row, to perform write operations. The
following code snippet copies i.e. writes variable x to specific \arch\  row and
column.} 


\begin{figure}[!htp]
	\centering
	\includegraphics[width=.47\textwidth]{Figs/code_snippet_01.pdf}
	\caption{Code snippet-01. 
		\label{fig:code_snippet_01}}
	\vshrink{0.3}
\end{figure}

{The above code-snippet shows how a variable is declared in \arch\ to store some
data and later read back to some user defined variable. \textit{x} and
\textit{y} are two integer variables. \textit{scratch\_preset} function declares
and presets some cells in each row as scratch area for \arch\ operation,
beginning at column, \textit{c} and extending over \textit{scratch\_length}
number of cells in each row. Since these presets are dependent on the type of
operations to be performed, the preset values are determined during compilation
by the compiler.  Now, a compiler provided variable type \textit{int\_pm} is
declared which points to the \arch\ location where a particular data will be
stored e.g. \textit{x}. A compiler specific function \textit{write\_pm} writes
\textit{x} at a row, specified by \textit{r}, beginning at column \textit{c},
and returns a pointer to the new stored variable as \textit{x\_pm}. In other
words, \textit{x\_pm} now represents data \textit{'1234'} stored in \arch\.The
subsequent read operation, invoked by \textit{read\_pm} copies back i.e. reads
the stored data, pointed by variable \textit{x\_pm} and store it in \textit{y}.     

After the data is written in \arch\, they can be processed through a sequence of
high level instructions to perform \arch\ supported logical operations,
accessible through compiler specified functions. Since the \arch\ array performs
every operation in parallel across all rows in an array, the corresponding code
only needs to specify the sequential operations to be executed along one row.
The developer is required to specify the locations of the operands, i.e. source
\arch\ variables, in one row, the type of operation to be performed and the
destination location i.e. another \arch\ variable initialized with some data
through \textit{write\_pm} function. For example, to perform \textit{NAND}
operation between two operands, the generic format of the corresponding
high-level instruction is: \textit{NAND (src1,  src2, destination)}. If the
operands are multi-bit variables, then it will be dismantled into sequence of
1-bit NAND operations by the compiler before dispatching the computation
requests to \arch\}. 


\begin{figure}[!htp]
\centering
\includegraphics[width=.35\textwidth]{Figs/code_snippet_02.pdf}
\caption{Code snippet-02. 
	\label{fig:code_snippet_02}}
\vshrink{0.3}
\end{figure}

Figure~\ref{fig:code_snippet_02} {illustrates the code sequence which performs
  NAND operation on two operands i.e. two 8-bit variables \textit{a\_pm} and
  \textit{b\_pm}, beginning at locations 1 and 9 and write the result in a 8-bit
  variable \textit{c\_pm}, beginning at location 17. Writing a single
  computation performs $NAND$ operation between \textit{a\_pm} and
  \textit{b\_pm}, stored  in the same row, and all other pair of variables which
  shares the same column locations in other rows. In other words, declaring and
  storing data in aligned manner across all rows, and specifying computation for
  only one row, it is possible to perform the same computation across all rows.   
More complex operations e.g. addition can be performed in a similar way as well.
The API call, \textit{ADD\_pm(a, b, d)}, adds two 8-bit variables $a$ and $b$,
beginning at locations 1 and 9, and stores the 9-bit result in 16-bit variable
$d$, beginning at column 17. This function performs 1-bit additions, one at a
time in one row, on all 8-bits of the input variables using corresponding full
adder equivalent operations. Since the basic addition operation is achieved
through 1-bit adder operations, this 8-bit operation is basically a collection
of multiple 1-bit addition operations. Also, there are intermediate operations
such as carry generation and propagation, preset cells for computation etc.,
which are abstracted by the compiler. During compilation time, this 8-bit
addition is first divided into basic addition operations e.g. 1-bit addition,
and then the more basic functions corresponding to addition operations e.g.
\textit{MAJ5} are inserted in-between and scheduled to achieve multi-bit
addition.   

\textbf{ISA Extension:} The compiler generates the executables from the sequence
of high level SpinPM operations, which contains the micro-instructions for the
memory controller driving the in-place computation in \arch\. These
micro-instructions are an extended set of low level instructions which are
received, decoded and executed by the SpinPM memory controller (SMC). Step-2
shows the micro-instructions for \textit{NAND}, \textit{NOR} and \textit{MAJ3}
operations. These micro-instructions only contain the type of operation and the
columns to connect as inputs and outputs.  For example, \textit{nand c3, c1, c2}
specifies column \textit{c3} as the output and column \textit{c1} and
\textit{c2} as inputs.

\textbf{Operation Scheduling:} This part of software stack is responsible for
optimizing the \arch\ access, for both read/write operations and computations
(Step-3). Optimization involves, mostly, scheduling of operations to \arch\,  in
order to achieve maximum possible throughput with a given data layout,
constrained by inherent parallelism in stored data. For example, the compiler
determines which writes and presets are to be merged together in each row, in
order to minimize the number of memory accesses, to compensate for the overheads
due to peripheral overhead associated with each access. Moreover, compiler
optimization breaks more complex functions provided by the compiler interface
into basic \arch\  operations with the objective of minimizing the number of
operations and data transfer between cells within a row.

\textbf{SpinPM Memory Controller (SMC):}  The SMC (step-4) is the hardware
controller for the \arch\  substrate. It handles synchronization between the
substrate and the host, through a synchornous protocol. Since each basic
computation takes a specific amount of time, it is possible to define a timing
window within which a requested task is expected to complete. This period of
time considers the \arch\  computation time, peripheral overheads and the
scheduling overhead coming form SMC itself. After this period of time is
elapsed, beginning from the time when the request is sent to SMC, the next batch
of controller specific requests i.e. micro-instructions are sent from the host.
Inside SMC, there is an instruction buffer which stores the incoming batch of
requests and stores them until they are issued to the substrate. Upon receving
the requests, SMC decodes the instructions to understand the type of operation
to be performed e.g. \textit{NAND} and selects the appropriate voltage level
before sending the control signals to the arrays with the column numbers to
operate on. This decoding process provides an opportunity to pipeline the
instruction decoding phase in SMC, to reduce the timing overhead due to SMC
operation.             }

}

}

\subsection{Practical Considerations}
\label{sec:overhead}
\noindent{\bf Array Size:}
The maximum row width (i.e., the max. number of columns) per \arch\ array
depends on the gate voltage $V_{gate}$ (Section~\ref{sec:fuse}), the interconnect
material for LL (which connects the input and output cells together in forming a
gate), as well as the technology node. We conduct the following
experiment to determine the max. row width: We consider a two-input, one output
\arch\ gate which has the input cells and the output cell located in adjacent
columns. In each experiment, we shift the output cell further away from the
input cells, by one cell at a time.  
The process continues until we reach the terminating condition,
which is when the current through the output cell falls below the
required critical switching current for the most conservative input cell
resistance states. 
Assuming copper interconnect segments of 160nm for LL, for representative \arch\
gates used in pattern matching, this analysis renders approximately 2K cells per
row at 22nm, where the latency overhead induced by this max. distance
computation barely reaches 1.7\% of the switching time of the MTJ (assuming a
near-term technology, as detailed in Section~\ref{sec:eval_setup}). 
\ignore{
{To determine the maximum row width for \arch, in
order to validate the row width considered during application mapping process, a
first order analysis is performed. As an example case, a two-input, one output
gate is considered which has the input cells located in adjacent columns, and
the output cell is shifted from the next column to columns which are further
away from the input cells. The process continues until the terminating condition
is reached which is when the current through the output cell falls below the
required critical switching current for the most conservative input cell
resistance states. 
Assuming \textit{LL} is made of copper interconnect segments which are 160nm in
length, connecting MTJ cells along a row, at 22nm technology node the maximum
row width is \~2000 cells. The fixed bias voltage considered here is 1.0V. The
corresponding delay is \textit{0.05 ns} which is 1.7\% of the switching latency
of near-term MTJ. The maximum row length is dependent on the bias voltage used
in gate configurations, the interconnect material, as well as the technology
node used.} 
}

\vshrink{-0.1}
\noindent {\bf Array Periphery:} 
Peripheral overheads, mainly induced by addressing and control operations, can play a vital
role in determining the pattern matching throughput. 
Accordingly, throughout the evaluation, we consider the time and energy overhead of
peripheral circuitry including row decoders, multiplexers, and sense amplifiers. 
For memory read and write operations a \arch\ array is not any different than a
standard STT-MRAM array, hence we model periphery after the standard STT-MRAM.
During computation, however, as all rows operate in parallel, row decoder
overhead does not apply (which we conservatively keep). The periphery during computation rather becomes similar to the periphery of Pinatubo~\cite{li2016pinatubo}, an alternative spintronic PIM substrate (although \arch\ computation relies on a different mechanism, totally excluding sense amplifier involvement during computation contrary to Pinatubo).
Even during computation where all rows are active, the current draw in an \arch\ array remains relatively modest.
For example, using projections for long-term MTJ
devices (as detailed in Section~\ref{sec:eval_setup}), a 128MB array would still consume considerably less current than a
DDR3 SDRAM write operation \cite{micron}.

\noindent {\bf Preset Overhead:} Each logic operation requires the output 
to be (pre)set to a predefined value. Computation is row parallel, i.e., in all rows, the output cell resides in the very same column. Accordingly, before firing row-parallel computation, the corresponding column where the output cells reside should be preset. 
To this end, we can use a 
``gang'' preset, which presets all cells in the output column simultaneously. The alternative is relying on the standard write operation, which can preset (columns in) one row at a time. Gang preset by definition is much faster than standard write based preset.
The gang preset is equivalent to a parallel COPY operation -- where all rows compute in parallel and where the output cells are all in the respective column subject to gang preset. Hence, the discussion about the periphery overhead during row-parallel computation directly applies here, and the current draw remains to be modest. For standard write based preset, on the other, the current draw is much less: As one gate can be actively computing in a row at a time, only one cell needs to be preset per row, and all rows are preset one after another. 

\ignore{
If the preset is performed along a column, which is basically a parallel \textit{copy} operation in all bits in a column, for a 10K-row array, it would take 1 A (for near-term MTJ) and 39.5 mA (for projected long-term MTJ) current, and 15.735 mJ (near-term) and 0.0695 mJ (long-term) energy. The latency for each technology is roughly equal to the specified switching latency for that technology. If standard write mechanism is selected instead, the energy required to preset exactly the same amount of bits in a column will be the same as the previous case, however this time, the current required is dictated by how many bits are written in one \textit{row} at time. Since the large arrays are created with rectangular tiles (subarrays), the actual current drawn, within a subarray, by a gang preset operation is 50 mA and 1.975 mA for near-term MTJ and future-term MTJ respecetively.             
}

\ignore{
for the performance of any kernel mapped in \arch. Peripheral
circuitry such as row decoder, mux decoder, sense amplifier etc. are considered
in evaluations of performance of mapped applications, to understand the impact
of such overheads. The read and write mechanisms in \arch are similar to that in
a standard STT-MRAM structure, and hence all the peripheral overheads are added
in estimation of energy and latency of SpinCM read and write operations. During
computation, however, all rows process data in parallel, and hence the row
decoder overhead is considered for all rows, although in real design the
overhead will be smaller.}

\ignore{

\subsection{Wiring Overheads}
\label{sec:overhead}

\subsection{Impact of Process Variation}
\label{sec:process_var}
\noindent We conclude the evaluation with a discussion on the impact of process variation, which, due to imperfections in manufacturing technology, may result in significant deviation in device parameters from their expected values. Both access transistors and the MTJ in an \arch\ cell are subject  to process variation.
Since access transistors are fabricated using the relatively more mature CMOS technology, the effect of process variation is far less dominating than what was in it's initial years. Being a relatively new technology, MTJ devices are more susceptible to process variation, which directly affects critical parameters such as switching current and switching latency. However, as MTJ technology matures, it is likely that it too will be able to reduce the impact of process variation.

One concern is variation in critical switching current, which can directly translate into variation in bias voltages on bitlines, i.e., $V_{gate}$, which determines the gate type. 
However, different {\arch} gates featuring close $V_{gate}$ values (and hence may be subject to this type of variation) are usually distinguished either by a different value of the preset or a different number of inputs, which makes it
 unlikely that the gate functions would overlap with each other as a result of variation. 
We validated this observation
assuming a variation in switching current by $\pm5\%$,$\pm10\%$ and $\pm20\%$, respectively, for all evaluated gates implemented in the \arch\ array.
\ignore{
the voltage range for 
\textit{MAJ5} gate is seen to have voltage range widened from 0.61-0.62V to 0.48-0.50V when process variation increases from 0\% to -20\%. With 0\% variation, \textit{MAJ5} has voltage range overlapping with that of $V_{TH}$. However, as the variation becomes -20\%, the voltage ranges do not overlap anymore.   
}

\ignore{
	\begin{table*}[!htp]
		\caption{Change in gate voltage range due to process variation.}
		\centering
		\begin{tabular}{|l|l|l|l|l|l|l|l|}
			\hline
			\textbf{var.} & 0\% & +5\% & -5\% & +10\% & -10\% & +20\% & -20\% \\
			\hline
			$NOT$ & 0.84-1.3 & 0.88-1.32 & 0.8-1.2  &   0.92-1.4 & 0.76-1.13 &1.0-1.5 & 0.67-1.00\\
			\hline	
			$NOR$ & 0.68-0.74 & 0.72-0.78 &  0.65-0.71 &  0.75-0.82 & 0.61-0.67 &0.82-0.89& 0.55-0.6\\
			\hline	
			$MAJ3$ & 0.65-0.69 &0.69-0.73 &  0.62-0.66 &  0.72-0.76 & 0.59-0.62 & 0.79-0.83& 0.52-0.56\\
			\hline		
			$MAJ5$ & 0.61-0.62 &0.64-0.65 & 0.58-0.59& 0.67-0.68 & 0.55-0.56 &0.73-0.74 & 0.48-0.50 \\
			\hline		
			$V_{TH}$ &0.62-0.63 &0.65-0.67 &  0.56-0.6&  0.68-0.70 & 0.55-0.57 &0.74-0.76 & 0.67-1.00\\
			\hline		
		\end{tabular}
		\label{tbl:process_var}
	\end{table*}      
	
	}   

}

\section{Evaluation Setup}
\label{sec:eval_setup}
\noindent {\bf Technology Parameters:} Table~\ref{tbl:mtj_param} provides technology
parameters for a representative near-term and projected long-term MTJ based
implementation.
The critical current $I_{crit}$ refers to an MTJ switching probability of 50\%,
which would incur a high write error rate (WER). To compensate, when deriving
gate latency and energy values, we conservatively assume a 2$\times$ (5$\times$)
larger $I_{crit}$ for the near (long) term MTJ technology.
We model access transistors after 22nm (HP) PTM~\cite{ptm}.

\ignore{
The subarray size was 512*512. Total STTMRAM size 128 KB. 

The bitline latency: 0.051189 ns (current), 0.063191 ns (future), bitline energy: 150 fJ (current and future). 

The write latency: 3.546 ns (current),  1.55 ns (future)

The read latency: 0.71 ns (current), 0.72 ns (future).

The write energy: 9028 fJ (current),  9027.00 fJ (future)

The read energy: 150.00 fJ (current and future).

technology node for transistors: 22 nm (HP). 
}

\vshrink{0.2}
\begin{table}[h]
	\caption{Technology Parameters.}
	\vshrink{0.1}
	\centering
	\resizebox{0.5\textwidth}{!}{
	\begin{tabular}{l|l|l}
		\textbf{} & \textbf{Near-term} & \textbf{Long-term}\\
		\hline
		\hline
		MTJ Type & Interfacial PMTJ  & Interfacial PMTJ\\
		MTJ Diameter (nm) & 45 & 10\\
		TMR (\%) &  133~\cite{r40} & 500\\
		RA Product ($\Omega\mu{m}^2$)&  5 & 1~\cite{r41}\\
		Critical Current $I_{crit}$ ($\mu{A}$)  & 100 & 3.95\\
		Switching Latency ($ns$) &  3~\cite{r42} & 1~\cite{r40}\\
		$R_P$ ($K\Omega$) &  3.15 & 12.7 \\
		$R_{AP}$ ($K\Omega$) &  7.34 & 76.39\\
		\hline \hline		
		Write Latency (ns)& 3.65 &  1.72 \\
		Read Latency (ns) & 1.21 & 1.24 \\
		Write Energy (pJ) & 0.36 & 0.308 \\
		Read Energy (pJ) & 0.83 & 0.78\\
		\hline \hline 
		$V_{INV}$ (V) & 0.84--1.3 &  0.23--0.48\\ 
		$V_{COPY}$ (V) & 0.84--1.3 &  0.23--0.48\\ 
		$V_{NOR}$ (V) & 0.68--0.74 & 0.20--0.22\\ 
		$V_{MAJ3}$ (V) & 0.65--0.69 &  0.20--0.21\\ 
		$V_{MAJ5}$ (V) & 0.61--0.62 &  0.19--0.20\\ 		
		$V_{TH}$ (V) & 0.62--0.63 &  0.19--0.20\\ 		
		\hline
	  \end{tabular}
	}
	\label{tbl:mtj_param}
	\vshrink{0.2}
  \end{table}

  \ignore{
\begin{table}[h]
	\caption{Peripheral Overheads.}
	\vshrink{0.1}
	\centering
	\resizebox{0.45\textwidth}{!}{
	\begin{tabular}{l|l|l}
		\textbf{} & \textbf{Near-term} & \textbf{Long-term}\\
		\hline
		\hline
		row decoder energy (pJ) & 0.11   & 0.11 \\
		mux decoder energy (pJ)& 0.17  & 0.15\\
		mux energy (pJ)& 0.069  &		0.046\\
		sense amplifier energy (pJ)& 0.256  & 0.256\\
		precharge energy	(pJ)& 0.22	& 0.22\\
		row decoder latency (ns)&  0.468867 & 0.468852\\
		mux latency (ns)& 0.004517  & 0.0012226		\\
		sense amplifier latency (ns)& 0.1  & 0.1\\
		precharge latency	(ns)& 0.525398	& 0.525398\\	
		charge latency		(ns)& 0.181084	& 0.248251\\
		\hline
	\end{tabular}
  }
	\label{tbl:overhead_val}
	\vshrink{0.2}
\end{table}		
}		

\vshrink{-0.1}
\noindent {\bf Simulation Infrastructure:}
We developed a step-accurate simulator in C++ to capture the throughput
performance and energy consumption of \arch\ based pattern matching as a
function of the technology parameters. 
We model the peripheral circuitry using NVSIM~\cite{nvsim} to extract 
the row decoder, mux, precharge, and sense amplifier induced energy and latency
overheads in \arch\ arrays used in the evaluation at 22nm.
%
Step-accurate simulation captures the overhead of each stage of pattern
matching:\\ 
(1) Write patterns on each row; \\
(2) Pre-set output cells (for comparison in match phase); \\
(3) Activate bitlines; \\
(4) Perform aligned comparison; \\
(5) Pre-set output cells (for similarity score computation phase); \\
(6) Activate bitlines; \\
(7) Compute score; \\
(8) Read-out score (optional).\\
Stages (2)-(4) are repeated for each bit of the pattern before moving to stage
(5), as an \arch\ row can only have one logic gate active at a time (i.e., we
can only perform one logic operation in a row at a time, but all rows can
compute that one operation simultaneously).  Finally, stages (2)-(8) are repeated
for each alignment (each at a different location of the reference fragment,
$loc$ per Algorithm \ref{algo1}), until the tails of the fragment and the pattern
meet. 
Due to row-level parallelism, the execution time of all of these stages in an array
is equivalent to the execution time in any row.                  
We derive energy consumption from this execution model, as well, where
the energy consumption of an entire array corresponds to
the sum of the energy consumption of each individual row in the array. 
Per array 
energy multiplied by the total number of arrays required to hold the reference
gives us the total energy consumption.      

\vshrink{-0.1}
\noindent {\bf Array Size \& Organization:}
For each benchmark, we simply stick to a straight-forward 2-bit representation for each character,
which yields the smallest possible array size. 
\ignore{
Each reference fragment (per row)
has 1K characters. Each array has 10K rows and 2K
cells, following the layout from Fig.~\ref{fig:layout}. 
We use a real human genome, NCBI36.54, from the 1000 genomes
project~\cite{1000genome} as the reference, and
4M 100-base character long real patterns from
SRR1153470~\cite{SRR1153470}.
}
It is evident that, depending on the pattern matching problem at hand, we might
need \arch\ arrays ranging from modest to very large in size. The thought
provoking issue here is how to deal with sufficiently large arrays as it might
restrict the design space, considering fabrication and circuit-level-design
related limitations. 
%
As an example, the proof-of-concept implementation 
requires 300 arrays of 10K rows and around 2K columns each for the string
matching case study from genomics.
This renders a total size of roughly 24Mb per array, which is not excessively
large. Still,
the fabrication technology might not be mature enough to synthesize such an
array.
Commercial MRAM manufacturers address this challenge 
by banking. 
For example, EverSpin~\cite{es} uses
8 banks in its 256 Mb ($32$Mb $\times 8$) MRAM product. Distributing
array capacity to banks helps satisfy the latency and energy requirement per
access, as well. 
For \arch\ based pattern matching, we too are inclined to use 
a hierarchy of banks, to enhance scalability.
While a clever data layout, operation scheduling and parallel activation
of banks can mask the time overhead, the energy
and area overhead would be largely due to replication of control hardware across
banks.    
The most straight-forward option for 
banked \arch\ would be to treat each bank simply as an
individual array which would map even shorter fragments of the reference to
patterns from the
input pattern dataset. 

\ignore{

This tiled architecture affects the overall data layout in various ways. To
illustrate the effects, we consider a generic pattern matching application. The
input patterns are written to different rows (depending on prior knowledge of
the possible match) to find a match to the already written reference patterns in
these rows. Since all or a number of rows of CRAM have to contain a segment of
reference pattern database and a copy of the input pattern, dividing an array
into tiles would mean some tiles will have only reference pattern, some will
have only input pattern, and the rest will have parts of reference and input
patterns. Since the input pattern has to be matched along whole reference
pattern segment i.e. across same row in different tiles, it is obvious that such
data layout may not be the best idea for such a mapping task. It would incur
data communication overhead (between different tiles, during computation) on top
of compute latency and energy. Fortunately, there is no actual transfer of data
in CRAM. Even if there is a scenario where data has to move from one part of the
array to the other, logic functions such as buffer is able to do it. The only
architectural feature needed in this case is a way to connect the LL of
corresponding rows in all tiles in x-dimension, and possibly selected tiles in
neighboring rows (across y-dimension)- in case of inter-row
computation/communication. However, there is an associated cost of this feature.
If the tiles in consideration are far apart, the resulting wire length
connecting the LLs of those tiles would pose limitation in terms of energy and
latency of the configured CRAM gates. As an example, if there is a logic gate
with two input cells in one tile and the output cell in another, and the tiles
are far apart in a large array, the LL wire connecting those tiles will have
larger resistance and parasitic capacitance. A possible solution to this problem
is to use a hierarchical structure involving local and global LLs to reduce
impacts resulting from those long wires. Another solution could be to use a hub
like switching node where the corresponding rows of each tile will be connected.
This way, the effective length of wire connecting LLs of to rows are same in all
cases, thereby eliminating asymmetry in latency of CRAM logic execution. The
design complexity associated with such a complex interconnection network might
be a challenge. 

\begin{figure}[!]	
	\centering
	\includegraphics[width=0.47\textwidth]{Figs/tiled_cram_March2.png}
	\caption{A tiled CRAM architecture}
	\label{fig_tiled_cram}
\end{figure}

}

\vshrink{-0.1}
\noindent {\bf Benchmarks:}
\label{sec:bench}
\noindent We evaluate \arch\ using four pattern matching applications (which
also include common computational kernels for pattern matching such as bit count), besides the running example of DNA sequence
alignment throughout the paper. 
Table~\ref{tbl:bench} tabulates these applications along with the corresponding problem sizes. 

\ignore{
{To corroborate the claim that {\arch} is highly efficient in processing
pattern matching workloads, four pattern matching applications are mapped to
{\arch} and evaluated.} Table~\ref{tbl:bench} \hl{shows the benchmark
applications and input datasets used in evaluation.} 
}

\vshrink{.1}
\begin{table}[h]
	\caption{Benchmark Applications.}
	\vshrink{0.3}
	\centering
	\resizebox{0.45\textwidth}{!}{
		\begin{tabular}{|l|l|l|l|l|l|}
			\hline
			\textbf{Benchmark} & \textbf{Reference/Problem Size} & \textbf{Pattern
				Length} & \textbf{Array Size}\\
			\hline
			\hline
			DNA & 3G char.  & 100 char. & 512$\times$512\\
			\hline
			Bit count & 1000000 32-bit vectors & 1-bit & 512$\times$512\\
			\hline
			String Matching & 10396542 words & 10 char. string & 512$\times$512\\
			\hline
			Rivest Cipher 4 & 10396542 words  & 248 bit & 1024$\times$1024 \\
			\hline
			Word count & 1471016 words & 32 bits & 512$\times$512\\
			\hline
			\hline
		\end{tabular}
	}
	\label{tbl:bench}
	\vshrink{0.2}
\end{table} 
 

\noindent {\em DNA sequence alignment (DNA)}
is our running case study throughout the paper.
We use a real human genome, NCBI36.54, from the 1000 genomes
project~\cite{1000genome} as the reference, and
4M 100-base character long real patterns from
SRR1153470~\cite{SRR1153470}.

\noindent {\em Bit count (BC)}~\cite{MiBench} counts the number of ones in a
set of vectors of fixed length. The counting consists of only
addition of bits in the vectors and then adding all individual counts. The input
vectors are mapped to the rows of {\arch} such that bit counting is performed in
parallel.   

\noindent {\em String Match (SM)}~\cite{phoenix} matches a search string
with a pre-stored reference string to identify the part of the reference
string of the highest or lowest similarity. Space separated string
segments and the search substring (which forms the pattern) itself are mapped to {\arch} rows such that
all searches are
performed in parallel.     

\noindent {\em Rivest Cipher 4 (RC4)} is a popular stream cipher. Upon generating
a cipher key, i.e., a string, it performs bitwise XOR on
the cipher key and the text to cipher. The same key is used to decipher the
text, as well. Segments of input text and the cipher key are mapped to {\arch} rows.

\noindent {\em Word Count (WC)}~\cite{phoenix} counts the number of occurrence of
specific words in an input text file, through word matching. The
words are mapped to {\arch} rows along with search words, and the word matching
in each row is executed concurrently.

\vshrink{-0.1}
\noindent {\bf Baselines for comparison:}
\label{sec:hmc}
\noindent{\em GPU Baseline:} To quantify by how much a {\arch} based
implementation of DNA sequence alignment provides improvement, we used a GPU
implementation of the commonly used BWA algorithm~\cite{li2008soap}. 
We use the very same reference and input pattern pool for the GPU baseline and
{\arch} mapped pattern matching application. Further, in order for the
comparison to be fair, we only take the pattern matching portion of the GPU
baseline into consideration (Section \ref{sec:patt}). 



\noindent{\em Near-Memory-Processing (NMP) Baseline:}
For throughput and energy characterization for near memory
processing based pattern matching, we use an HMC model based on
published data~\cite{hmc}. HMC power and latency models have contributions from
three components: memory and logic layers, and communication links. To favor
the NMP baseline, we ignore the power required to navigate the global wires
between the memory controller and the logic layer, and intermediate routing
elements. For logic layer, we consider single issue in-order cores, modeled
after ARM Cortex A5~\cite{A5_ref} with 1GHz clock and 32KB instruction and
data caches. The cores have a peak power rating of 80mW, with dynamic power
varying between 30mW and 60mW~\cite{pugsley2014ndc}. We first consider a total
of 64 cores to provide parallel processing, which renders a total peak power of
5.12W.  
\ignore{ However, to further favor the baseline, we assume a dynamic
power of 20 mW only per core.} 
For communication, we assume an HMC-like configuration with
four communication links operating at their peak frequency of 160 GB/s. To
derive the throughput performance, we use the same reference and input patterns
to profile 
each benchmark. We then use the instruction and
memory traces to calculate the throughput. We validated this model through CasHMC~\cite{casHMC} simulations. For reference, we also include a
hypothetical NMP variant which includes 128 cores in the logic layer,
and incurs zero memory overhead.  

\ignore{ The execution latency of different
  instructions are assumed to be smaller than published data. For instance, the
  execution latency for integer and floating-point instructions are reported to
  be 3 and 5 cycles respectively~\cite{azizi2010energy}, whereas we assume 1 ns
  latency for both type of instructions, in order to favor NMP model.  }


\ignore{
Recall that we
do not consider various components of the communication link, DRAM layer energy
overhead etc. For instance, considering 5.76 W of power consumed by the SerDes
circuit~\cite{jeddeloh2012hybrid} alone, the compute efficiency comes down to
3.75 seq/mJ. Also, the execution latency of different instructions are assumed
to be smaller than published data. For instance, the execution latency for
integer and floating-point instructions are reported to be 3 and 5 cycles
respectively~\cite{azizi2010energy}, whereas we assume 1 ns latency for both
type of instructions, in order to favor NMP model.
}

\ignore{
\begin{table*}[!]
	\caption{Throughput Comparison (with pattern length of 100 bp).}
	\centering
	\begin{tabular}{|l|l|l|l|}
		\hline
		\textbf{Parameter} & \textbf{SOAP3-dp} & \textbf{SpinCM++} & \textbf{SpinCM++(proj.)}\\
		\hline
		Throughput (Kseq/sec) & 104.61  & 9957.12  &  21743.3 \\
		\hline	
	    Power (W) & 86.302 (avg.) & 135.82  & 298.06 \\
		\hline	
		Compute efficiency (seq/mJ) & 1.21 & 73.3  & 72.94 \\
		\hline		
	\end{tabular}
	\label{tbl:comparison_performance}
\end{table*}

\begin{table}[ht]
	\caption{Pattern Matching on Near Memory Processing Substrate\label{tbl:nm_eval}}
	\centering
		\begin{tabular}{|l|l|}
			\hline 
			{\bf Performance metric} & {\bf Value} \\
			\hline
			Power (W) &  1.28 \\
			\hline
			Throughput (Kseq/s) & 27.28  \\
			\hline
			Compute Efficiency (Seq/mJ) & 21.31 \\
			\hline
		\end{tabular}
\end{table}

}


\ignore{
\subsection{System Modeling}
Through the pseudo-code, we show how in-place computation progresses in one row.
Since all the rows executes the same computational steps in parallel,
understanding computation from the perspective of one row is enough. We can
divide the compute tasks in to two phases. The first phase, \textit{match
phase}, aligns the pattern to the reference segment in the same row. The next
step is responsible for counting how many matches (or mismatches, depending on
the algorithm) has the match phase generated during a particular iteration. This
phase is called \textit{score compute phase}. Algorithm \ref{algo1} illustrates
how the sequence of operations are carried out in a row. The computation begins
with aligning the heads of reference segment and the pattern (i.e. from string
location 0), and iterates continuously until the pattern reaches the tail of the
reference segment. At each iteration,  the pattern is matched to reference
segment and the resultant output string contains 0s and 1s (0 for a match and 1
for a mismatch). The number of 1s in the string is a measure of how similar is
the pattern to the reference segment at that particular location. This is done
using a network of adder, forming a reduction network, counting the number of 1s
in the string. When the reduction of match string is done, the location is
incremented by 1 i.e. the matching is shifted along the reference sequence. Each
of the operations are preceded by presetting of output cells to ensure that the
outputs do not have any noise from previous iterations.

To understand, how the system performs, we assumed two separate models for
	   energy and latency estimations. However, these two models are similar in
	   the sense that they follow the same algorithm.     
       
\subsubsection{Latency Model}
\label{anamodel_latency} Figure ~\ref{fig_latency_model} shows the timing
diagram corresponding to the compute phases from Algorithm ~\ref{algo1}. The
write bits refers to the writing of pattern bits on each row of each array. The
latency associated with bitline activation is considered here as well. Once the
bits are written to an entire array or system of arrays, the \textit{match
phase} begins. The operations begins with matching the heads of pattern and the
corresponding reference segments. The \textit{match phase} is an iterative
process which loops over the entire length of the reference segment on a
particular rows. Before performing the match operations, the corresponding
output bits are preset to the required state (0 or 1) and the bitlines
associated with the input and output cells are activated. Upon exiting this
loop, it has a string of 0's and 1's in the output cells. The bits are counted
with the help of a series of adders in reduction tree like approach. In each of
the adder steps in \textit{compute score phase}, the corresponding output cells
are preset and the bitlines are activated. When the \textit{score compute phase}
is complete, there is a window of wait time introduced to allow for the scores
to be read from the rows of an array (buffer read). The whole process iterates
over the entire length of the reference segments, with the head of pattern
aligned with the next base character- starting from the head of the
corresponding reference segment. Once the pattern reaches the end location on
reference segment i.e. $reference\_segment\_length - pattern\_length$, the
similarity search process is complete. The latency (and throughput) of an array
is simple the latency of any row since all rows perform similarity search in
parallel.                  

\begin{figure*}[!]	
	\centering
	\includegraphics[width=0.95\textwidth]{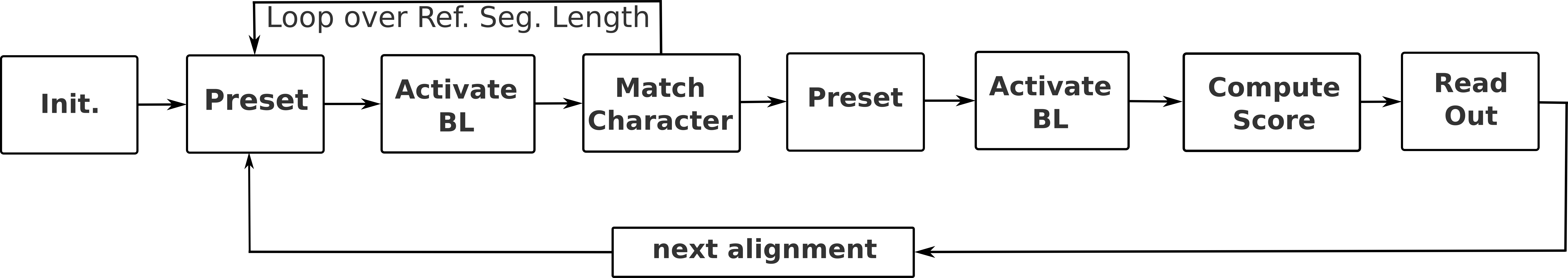}
	\caption{Process flow diagram}
	\label{fig_latency_model}
\end{figure*}

\subsubsection{Energy Model} \label{anamodel_energy} The energy model is derived
from the latency model, with the exception that the energy of an entire array
has contributions from individual rows in that array. The steps/operations shown
in Figure~\ref{fig_latency_model} is repeated for every row of an array, and the
aggregated energy contributes toward the total the energy of the design. This
energy multiplied by total number of arrays required to hold the reference
sequence is the total energy consumed by the designed system.      
}


\section{Evaluation}
\label{sec:eval}
\noindent 
We will start the evaluation with detailed throughput performance and energy
characterization, along with a sensitivity study, using {\em DNA} as a case study.
Specifically, we will consider two design points, which differ in how the
patterns (from the input
pattern pool) get assigned to rows for matching. In other words, how patterns are
{\em scheduled} for computation in the \arch\ array:
The first one is a {\em Naive} implementation, where we take one pattern and blindly copy it to
every row of all arrays to perform similarity search. 
The second implementation, on the other hand, features {\em Oracular} pattern
scheduling, which can avoid assigning a pattern to a row where a too dissimilar
(reference) fragment resides.
{\em Oracular} is straight-forward to implement by adding a pre-processing step, where hash-based filtering is not uncommon~\cite{kimgenome}. We will leave exploration of this rich design space to future work.
Any practical \arch\ implementation would fall somewhere in the spectrum
between these two extremes.

\noindent {\bf Naive Design ($Naive$):}
The caveat of this approach is that, since this design accepts
one pattern at a time and aligns it naively to all reference fragments, the
overhead of redundant computation is very large. 
Moreover, as a single pattern is matched to the entire reference, across all
arrays, at a time, 
the apparent
serialization 
hurts the throughput of the system, in terms of the number of patterns matched
per second. In the following, we will refer to the number of patterns matched
per second as {\em match rate}.


\noindent {\bf Oracular Pattern Scheduling ($Oracular$):}
The oracular scheduler resides between the input pattern pool and \arch, and
controls to which row in which array each pattern goes. $Oracular$ may still
feed a given pattern to multiple rows, in multiple arrays, however, does not
consider rows which carry a too dissimilar (reference) fragment.   
In other words, $Oracular$ directs patterns to rows and arrays in a way such
that achieving a high similarity score becomes more likely.
While $Oracular$ bases its pattern scheduling decisions on perfect information,
a practical implementation of this idea would incur 
the overhead of gathering this information, i.e., extracting a schedule to keep
pattern matching confined to rows where a high similarity score is more likely.
In any case such smart scheduling 
of patterns benefits the throughput performance by reducing redundant
computation which eats from the energy budget. 

However, since all rows in a \arch\  array perform pattern matching (in
lock-step but) in parallel, before computation begins, we require that
all rows have their patterns ready.  Scheduling patterns takes time, which might
further affect the throughput performance of \arch, if we let the array sit
idly,
waiting for scheduling decisions to take place.
We can mask this overhead,
as
drawing pattern scheduling decisions for all the rows in
an array 
takes less time than 
writing patterns in the rows of that
array. This, in effect, would not introduce any timing overhead towards the system
throughput, although there is an energy overhead.
 
\subsection{Throughput Performance and 
	Energy Characterization}
\label{sec:char}
\begin{figure}[!]	
	\subfloat[Match Rate (patt/sec)]{
		\centering
		\includegraphics[width=0.23\textwidth]{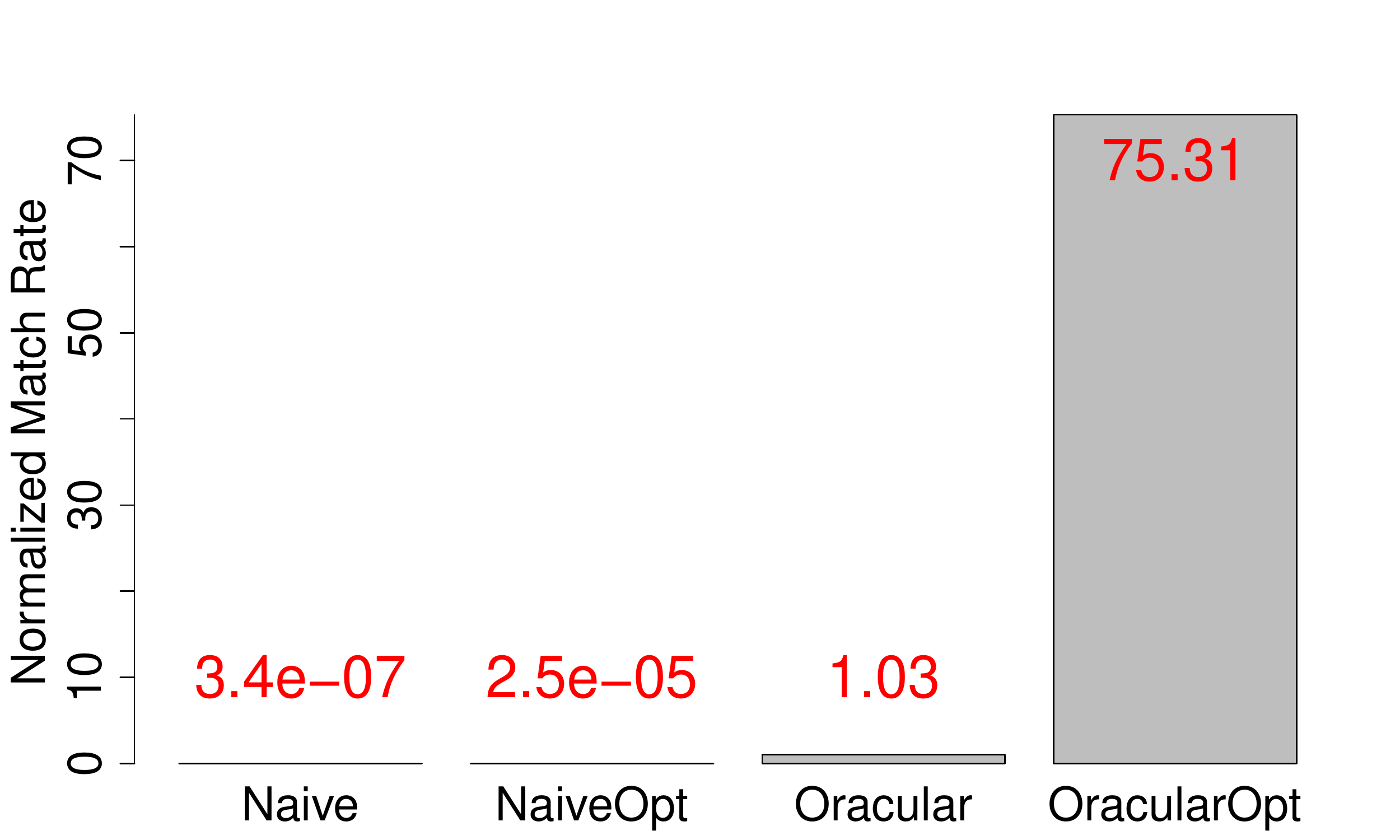}
		\label{fig_match_rate_norm_1}
	}
	\subfloat[Compute Eff. (patt/sec/mW)]{
		\centering
		\includegraphics[width=0.23\textwidth]{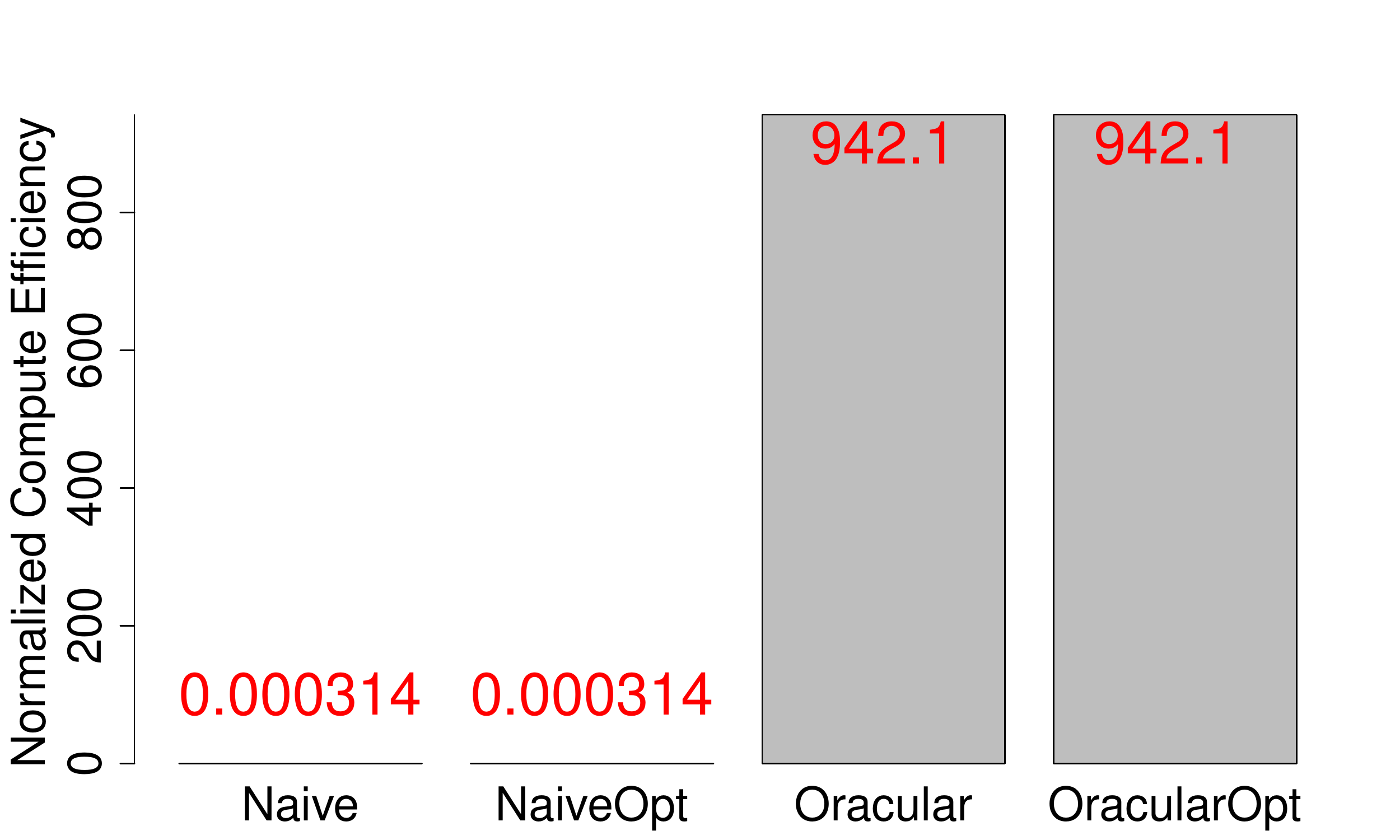}
		\label{fig_comp_eff_norm_1}
	}
	\caption{Performance and Energy Characterization.}
	\label{fig_normalized_figure01}
	\vshrink{0.5}	
\end{figure}

\noindent 
Fig.\ref{fig_normalized_figure01} shows the throughput performance and energy efficiency, normalized to GPU baseline, for $Naive$ and
$Oracular$, when processing 
a pool of 3M patterns. 
We use match rate (in terms of number of patterns processed per second) for
throughput; match rate per milliwatt, for energy efficiency.
$Naive$ 
yields very low throughput -- by mapping
each pattern to every row of each array at a time, and thereby increasing the
total execution time significantly. 
$Oracular$ pattern scheduling is very effective in eliminating this
inefficiency: we observe that the throughput performance w.r.t. $Naive$
increases by approx. close to an order of magnitude in this case.

To put these throughput values in context, we can look at the time required to
process the pool of 3M patterns, which is over 23215.3 hours, using 300 arrays
under $Naive$.
The fundamental limitation for $Naive$ is the redundancy in
 computation. 
 Since at a time, $Naive$ feeds only one
 pattern into all \arch\ arrays, the total time required to process the
 entire pool of patterns is higher. The effective throughput is limited by the
 time taken to align one pattern in one row. 
 On the other hand, $Oracular$ only takes 
 about $2.32$ hours for the same pool of patterns. This drastic change in
 runtime is due to feeding multiple patterns into \arch\ arrays at
 the same time. 

\begin{figure}[!]	
	\subfloat[Energy]{
		\centering
		\includegraphics[width=0.23\textwidth]{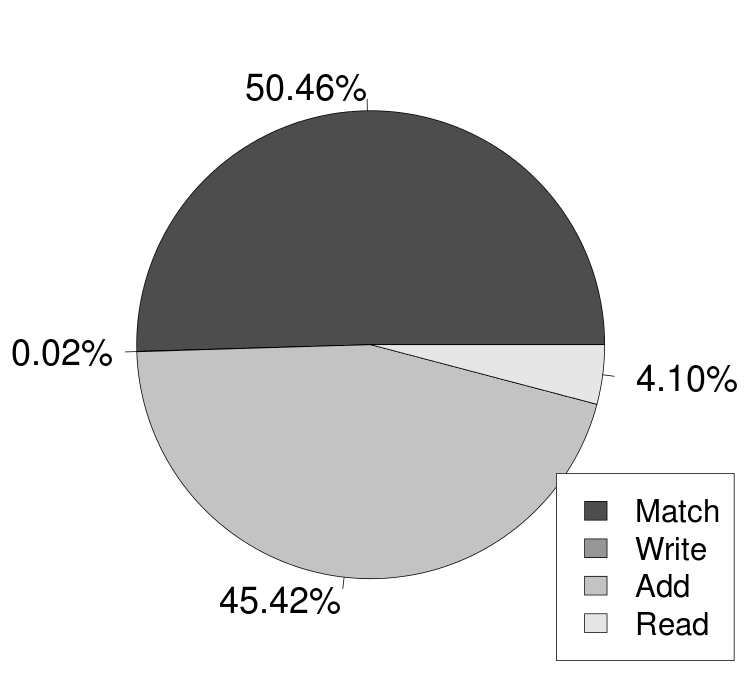}
		\label{fig_unoptimized_energy}
	}
	\subfloat[Latency]{
		\centering
		\includegraphics[width=0.23\textwidth]{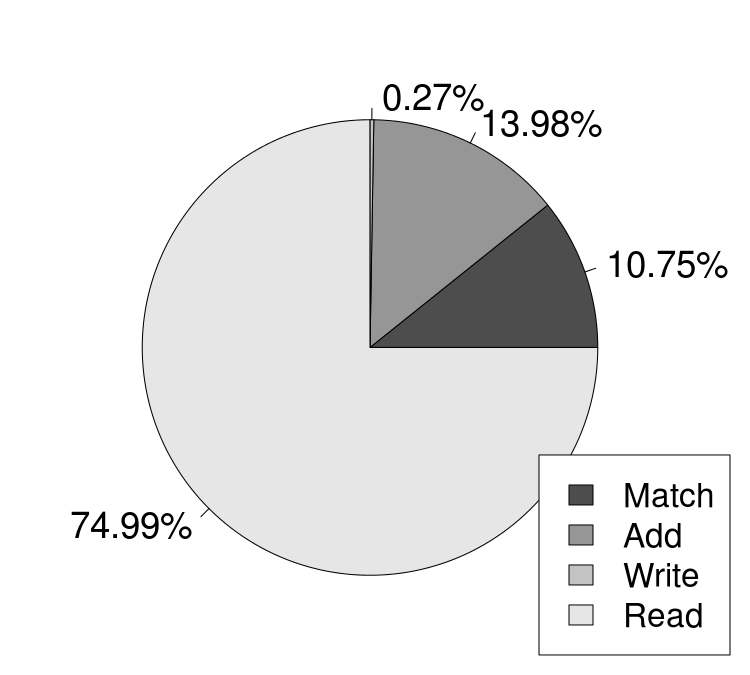}
		\label{fig_tech_unoptimized_latency}
	}
	\hfill
	\caption{Breakdown of energy and latency in computation.}
	\label{fig_unoptimized_breakdown}
	\vshrink{0.5}	
\end{figure}
It is fundamental to the understanding of the performance and energy
characterization to identify the
  individual contributions of actual computation stages  -- i.e., Stages(1)--(8) from
  Section \ref{sec:eval_setup}.
  Fig.\ref{fig_unoptimized_breakdown} shows the distribution of energy and
  latency components. The preset overheads are 43.86\% and 97.25\% in energy and
  latency, respectively, where the bit-line (BL) driver energy and latency overheads
  are <1\% and 2.7\% respectively.
  The
  breakdowns in Fig.\ref{fig_unoptimized_breakdown} do not contain preset and BL driver related overheads.
  Apart from these, we observe that  
  the majority of the energy (Fig.\ref{fig_unoptimized_energy}) is consumed by the match operations 
  and additions during similarity score computations. However, in case of
  latency (Fig.\ref{fig_tech_unoptimized_latency}), the dominant components change to
  read-outs of similarity scores (i..e, Stage (8)) and
 additions during similarity score computations. In case of both energy and latency,
  writes (i.e., Stage (1)) consume $<1$\% of the share. 
  
  This breakdown 
 clearly identifies preset overhead as the essential bottleneck.
  Also, 
  although the time required by the match
  and similarity score compute phases are not drastically different, the
  energy required by the similarity score compute phase is around twice of that
  of match phase. 
  Accordingly, we next look into preset and similarity score computation
  operations for optimization opportunities.
 

\noindent {\bf Optimized Designs ($NaiveOpt$, $OracularOpt$):}
As the reduction tree for addition (Fig.\ref{fig:data_op_score_comp}), which is at the core
of similarity score computations, already represents an efficient design, we
focus on optimizations to reduce the preset overhead. 
Since presets are inevitable for
logic operations, it is not possible to entirely get rid of them.
However, we can still hide preset latency through careful scheduling
of presets.

As presets do  not correspond to actual computation, $Naive$ and $Oracular$ simply perform them in
between computation.
The challenge comes from 
successive steps in computation using
the very same set of cells to implement logic functions. Instead of interrupting
computation to preset these cells every time a few computation steps are
completed, 
we can distribute such consecutive steps to different cells, using the scratch
area from Fig.\ref{fig:layout}, and preset them at once, before computation
starts.
We call the resulting designs $NaiveOpt$ and $OracularOpt$, respectively.
The \textit{NaiveOpt} and \textit{OracularOpt} bars in Figure~\ref{fig_match_rate_norm_1} and Figure~\ref{fig_comp_eff_norm_1} capture the resulting 
energy and throughput performance.  
We observe that, for each design option, energy consumption of the optimized
case is unchanged. This is because the optimization only changes the scheduling
of presets, where the total number of presets performed still remains the same. 
The throughput performance, on the other hand, skyrockets in both cases thanks to gang presets (Section~\ref{sec:overhead}).

\ignore{
\begin{table}[ht]
	\caption{Evaluation Results (SpinCM++). \label{tbl:eval_result_optimized}}
	\centering
	\resizebox{0.45\textwidth}{!}{
		\begin{tabular}{|l|l|}
			\hline 
			{\bf Performance metric (per array)} & {\bf Value} \\
			\hline
			Energy &  136.42 (mJ) \\
			\hline
			Case-1: Throughput (no filter) &  3.32 (seq./sec) \\
			\hline
          	Case-1: Compute Eff. (no filter) &  0.0073 (seq./mJ) \\
			\hline
            Case-2: Throughput (with filter) &  33.19 (Kseq./sec) \\
			\hline
            Case-2: Compute Eff. (with filter) &  73.3 (seq./mJ) \\
			\hline
		\end{tabular}
	}
\end{table}

Only the fraction of
preset latency is now reduced to 95.33\%. In terms of absolute reduction, from
Table~\ref{tbl:latency_optimization}, it is obvious that the reduction in preset
and BL driver latencies are significant. The preset latency, in this case, is
reduced 104.67 times relative to the unoptimized SpinCM. At the same time, since
the corresponding BL driver latency also decreases, the overall BL driver
latency contribution decreases as well.          
\begin{table}[ht]
	\caption{Effect on preset latency after optimization. \label{tbl:latency_optimization}}
	\centering
		\begin{tabular}{|l|l|l|}
			\hline 
			{\bf SpinCM } & {\bf Preset Latency (ms)} & {\bf BL Latency (ms)}\\
			\hline
			Unoptimized &  30063 & 433.13\\
			\hline
			Optimized &  287.23 & 3.29\\
			\hline
		\end{tabular}
\end{table}
}






\noindent{\bf Practical Considerations (Pattern Scheduling):}
\noindent The throughput we reported for $Oracular$ is
the theoretically achievable maximum.
How close a practical implementation can come to this strongly depends on the
actual values of the patterns, as well, which may or may not ease scheduling
decisions.  
Since each array keeps consecutive fragments of the 
reference,
it is always possible that patterns directed into a particular
array do not have any matches in any of the rows.
We may not always be able to eliminate such ill-schedules, depending on the pattern
values, where the incurred   
redundant computation would degrade performance.
\ignore{
Therefore
it helps to have some sort of filter which is capable of reordering the sequence
in a way so that the most of the rows in a particular array have different
patterns mapped to them. Since patterns are not processed on-line, it should be
possible. It is obvious from the previous discussion that having an ideal (no
overhead) and perfect scheduler enables us to achieve a high throughput.
}
The feasibility of any pattern scheduler is contingent upon the distribution
of the patterns, in terms of the rows in the arrays where the most similar
fragments reside.

\ignore{

\begin{figure}[!]	
\centering
	\includegraphics[width=0.47\textwidth]{Figs/histoarray1_march19.jpg}
	\caption{Pattern Histogram.}
	\label{fig:histo}
\vshrink{0.2}
\end{figure}

Figure~\ref{fig:histo} shows one such representative
histogram of row hits where only rows having index of 1000 to 2000 is shown for
better visibility. The trend for different range of rows in that array and
across multiple arrays are very similar.  

A practical pattern scheduling approach based on hashing would work as follows:
A \emph{k-mer} is the substring of a string having \emph{k} characters.
We can take an example reference, and extract
\emph{k-mers} starting at each possible location of the
reference.

see different patterns appearing in that
long string, some of which are repetitive even. We store the patterns, 

with the help of a hash function which renders a different integer number for
each different
\emph{k-mer}, and create a map of the rows (i.e. reference fragments) which have
\emph{k-mers} producing those values.

Once we have the map, the patterns are
input in batch and the first \emph{k} base characters of each pattern are fed to
the hash function to get the unique integer value which is then used to get the
row numbers in the array. We performed this analysis for multiple arrays (i.e.
$1000 \times 10000$ base characters for one array), at a time and observed the
distribution of row hits i.e. rows having number of possible matches with
different patterns. Figure~\ref{fig:histo} shows one such representative
histogram of row hits where only rows having index of 1000 to 2000 is shown for
better visibility. The trend for different range of rows in that array and
across multiple arrays are very similar.    

This motivational example clearly indicates 

It is obvious that the distribution of the patterns to the rows of an array
corresponds to no distinct pattern, meaning that there is opportunity to reorder
sequence in a way so that when patterns arrive at the arrays, they are more
evenly distributed throughout the array, thereby reducing the amount of penalty
due to unavailability of patterns for any row. To get the maximum performance
out of the system the scheduler has to be capable of reordering the patterns
accordingly to make sure that at one run, maximum number of rows have different
patterns from the pool of patterns.   

}

\subsection{Sensitivity Analysis}
\label{sec:sensit}
\begin{figure}[!]	
	\subfloat[Match rate (patt/sec)]{
		\centering
		\includegraphics[width=0.23\textwidth]{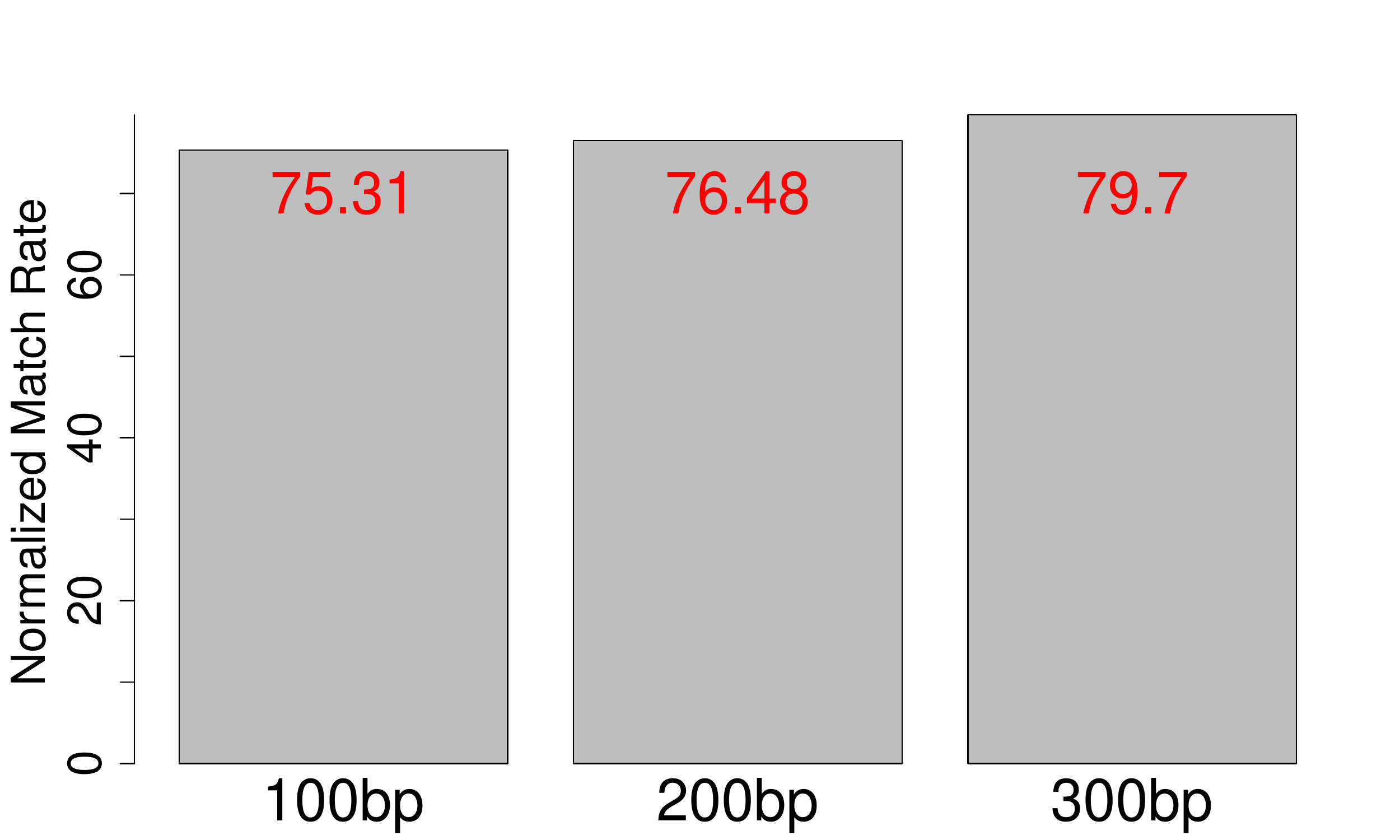}
		\label{fig_match_rate_norm}
	}
	\subfloat[Compute Eff. (patt/sec/mW)]{
		\centering
		\includegraphics[width=0.23\textwidth]{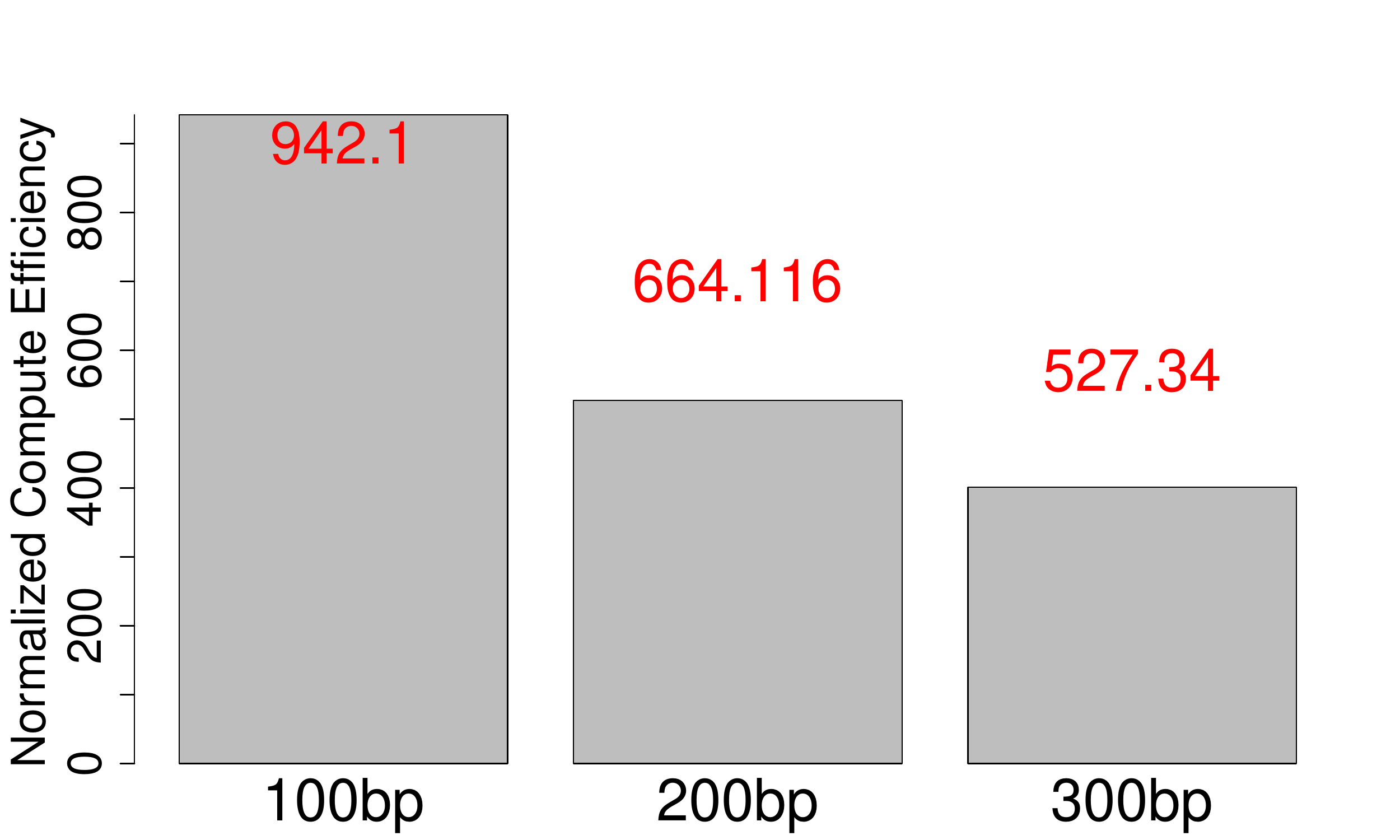}
		\label{fig_comp_eff_norm}
	}
	\caption{Sensitivity to pattern length for \textit{OracularOpt}.}
	\label{fig_normalized_eval_pattern_length}
	\vshrink{0.5}	
\end{figure}

\noindent{\bf Sensitivity to Pattern Length:}
Up until now, we have used
a pattern length of 100 characters. 
We will next examine the impact of pattern length on energy and throughput
characteristics. Without loss of generality, we confine the analysis to
$OracularOpt$.
For the purpose of design space
exploration, we experiment with pattern lengths of 200 and 300 characters, which
are representative values for the alignment of short DNA
sequences~\cite{illumina}. We keep the array structure the same, while the
reference length remains fixed by construction.
%
Fig.\ref{fig_normalized_eval_pattern_length} summarizes the outcome.
Understandably, with the pattern length
increasing, more computation becomes necessary to generate the similarity
scores in each row. However, this effect does not directly translate into
degraded performance: The throughput for
increasing pattern lengths remains close to the baseline throughput for 100-character patterns.
This is because the preset optimization is scalable.
Increasing pattern length translates into more scratch bits for presets, which
acts against throughput going down sharply. 
 %
Irrespective of the application domain, the maximum pattern length is actually
limited by technology constraints, since the required number of cells per row
also increases with increasing pattern length. 
We further observe that the compute efficiency (i.e., the match rate per mW) 
decreases due to
increases in computation per alignment, which is congruent with the intuition. 


\begin{figure}[!]	
	\subfloat[Match rate (patt/sec)]{
		\centering
		\includegraphics[width=0.23\textwidth]{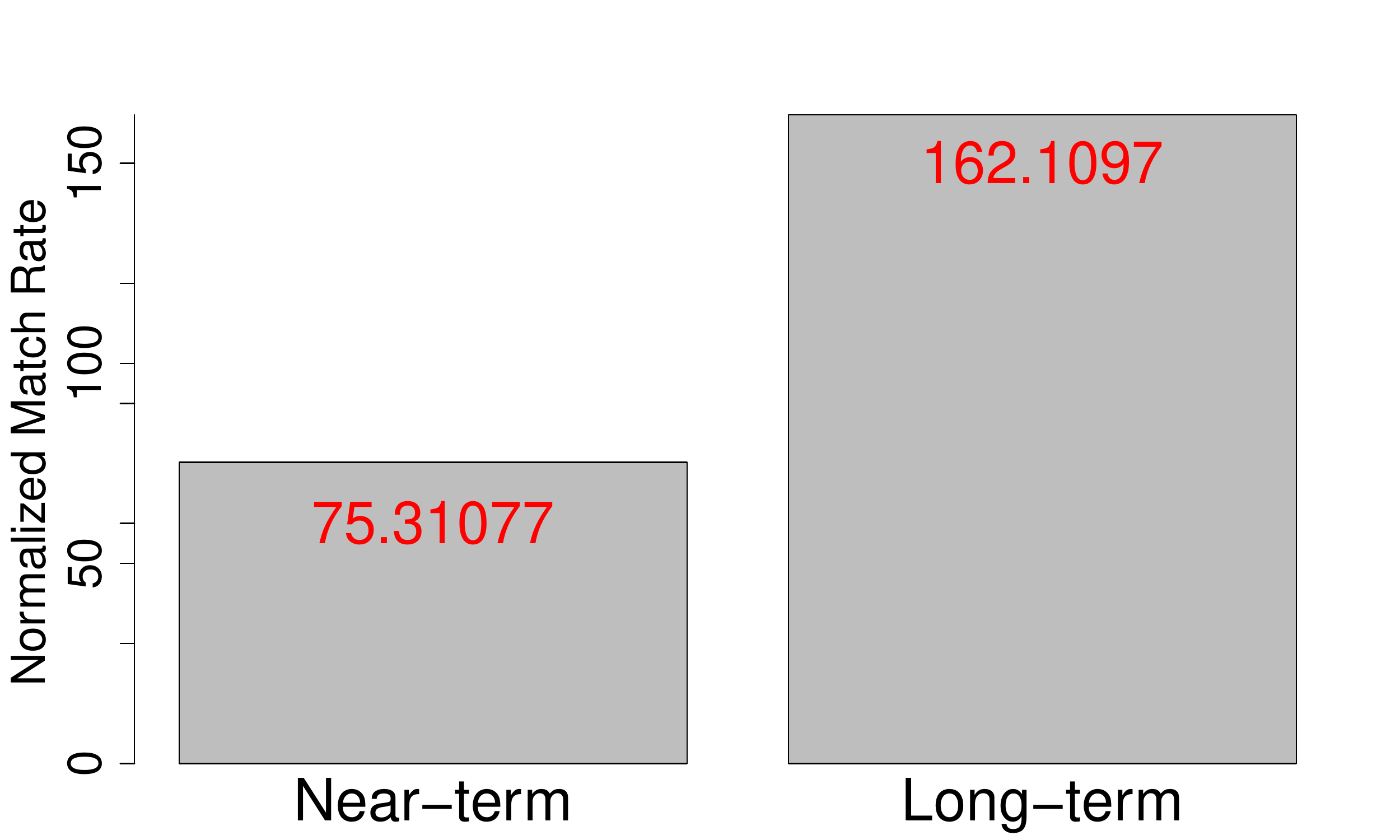}
		\label{fig_match_rate_norm_tech}
	}
	\subfloat[Compute Eff. (patt/sec/mW)]{
		\centering
		\includegraphics[width=0.23\textwidth]{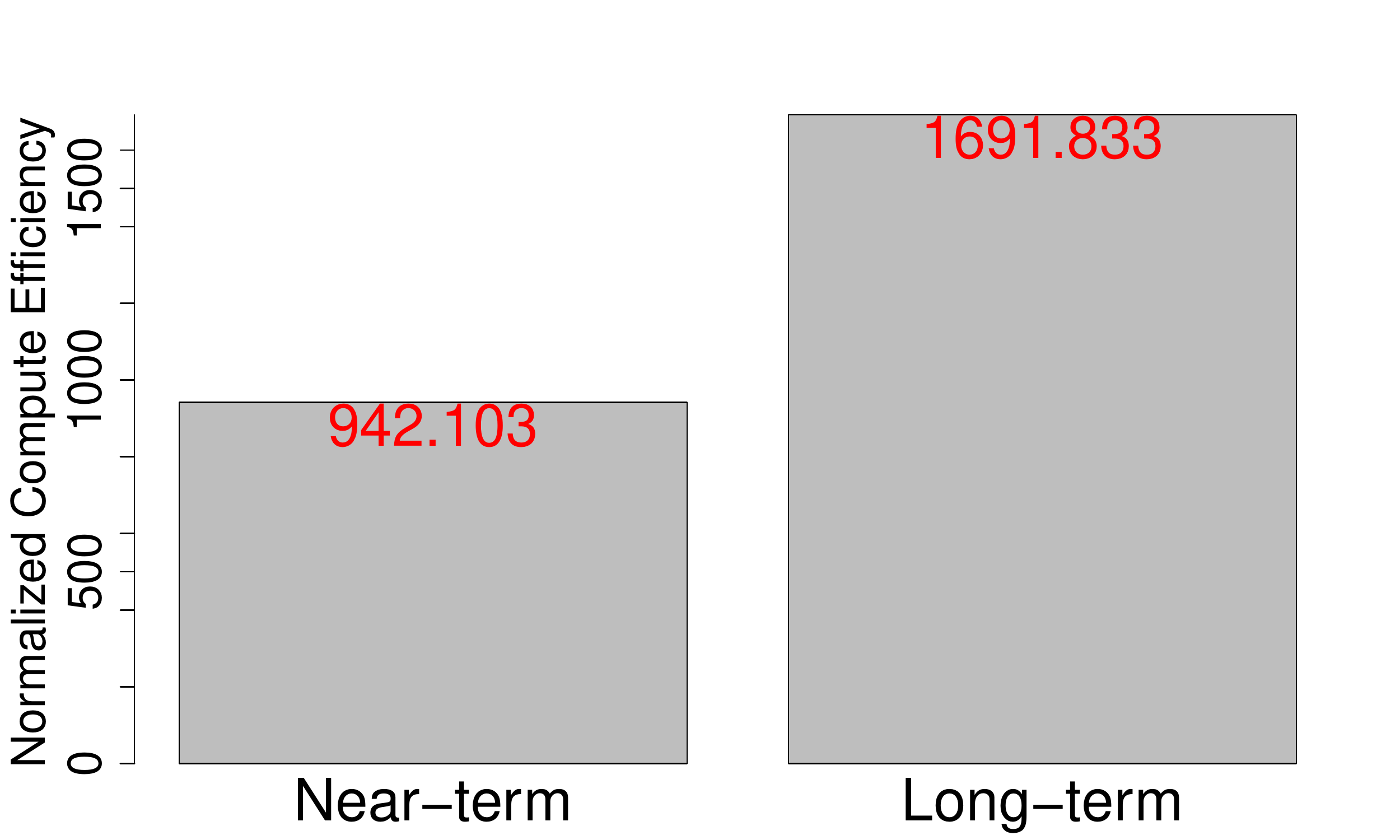}
		\label{fig_comp_eff_norm_tech}
	}
	\caption{Sensitivity to MTJ technology for \textit{OracularOpt}.}
	\label{fig_tech_trend_throughput_power}
\end{figure}

\noindent{\bf Sensitivity to MTJ Technology:}
MTJs have been able to meet technology trend estimations so far. We next consider the long-term
technology projections from Table~\ref{tbl:mtj_param} for the default, representative pattern length of 100. Building upon \textit{OracularOpt}, we will refer to this
design as \textit{OracularOptProj}.
As Fig.\ref{fig_tech_trend_throughput_power} indicates, 
a boost 
in match rate (i.e., throughput) and compute efficiency by approx. 2.15$\times$ becomes possible.


\subsection{\arch\ vs. NMP}
\label{sec:bench_eval}
\noindent In the following we 
characterize benchmark applications, in terms of {match rate} and {compute efficiency}, when mapped in \arch\ vs. two baselines: NMP and a hypothetical variant of NMP with no memory overhead ({NMP-Hyp}).  
\begin{figure}[!htp]	
	\centering
	\includegraphics[width=0.47\textwidth]{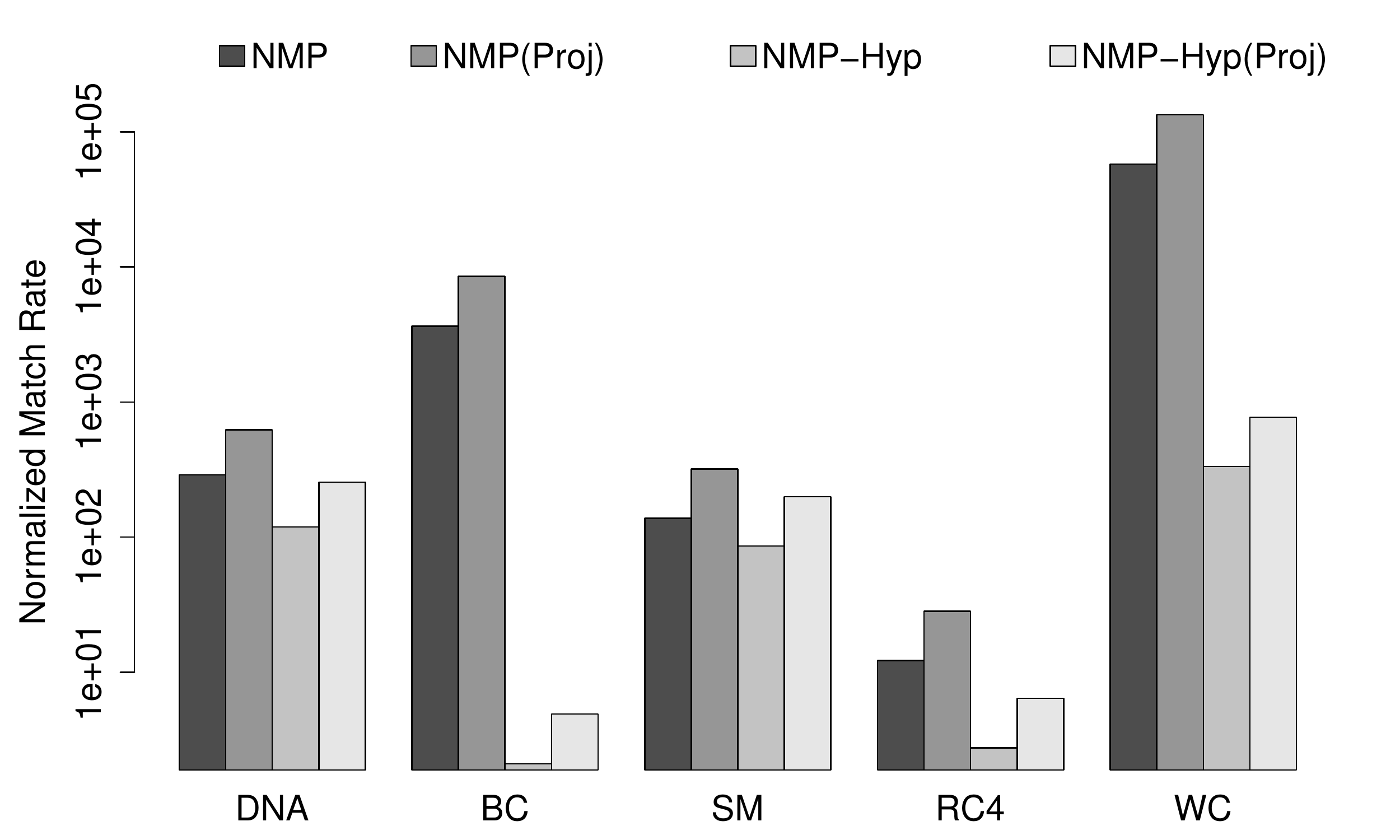}
	\caption{Normalized Match Rate (patt/sec) in Log scale.}
	\label{fig_patt_all_norm}
\end{figure}

\begin{figure}[!htp]	
	\centering
	\includegraphics[width=0.47\textwidth]{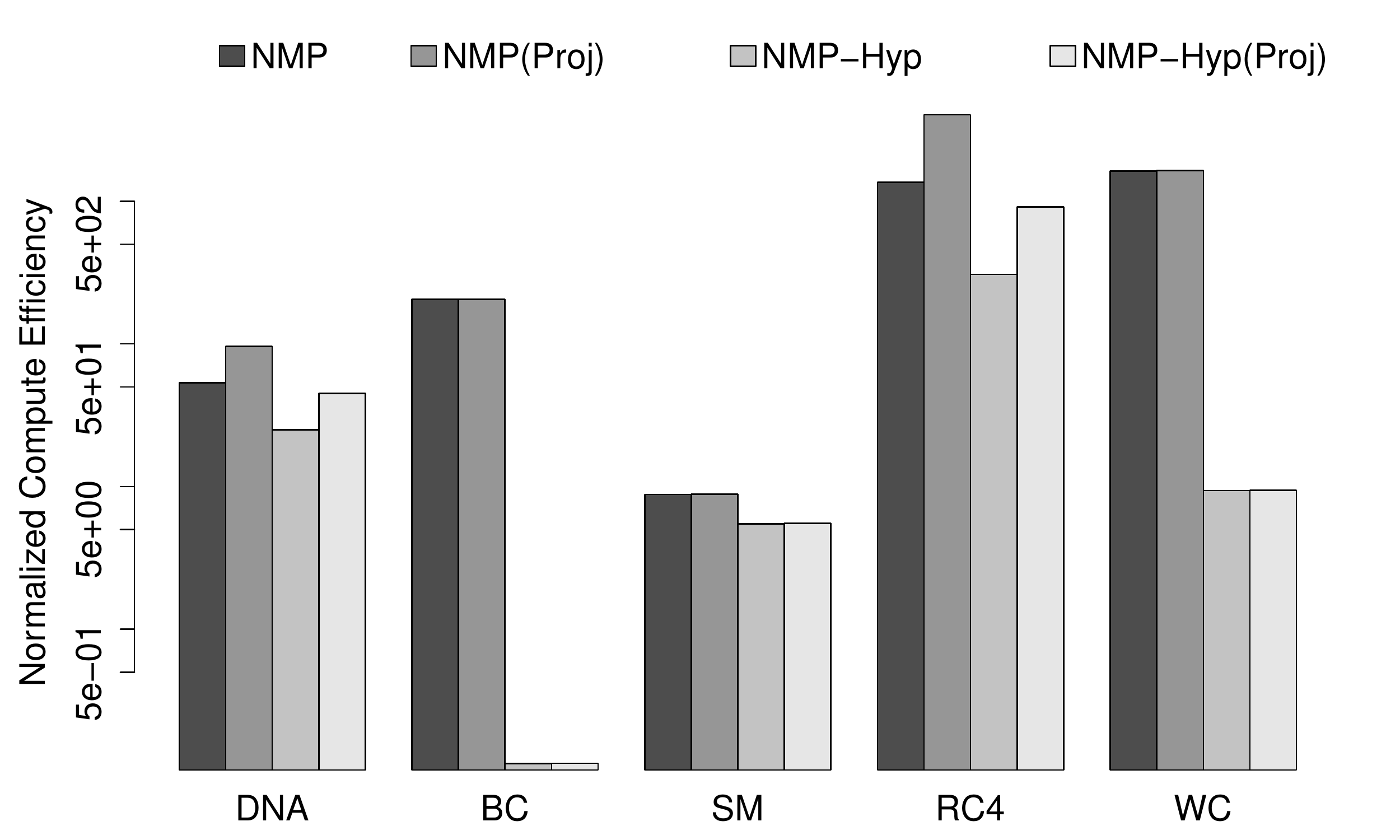}
	\caption{Normalized compute efficiency (patt/sec/mW) in Log scale.}
	\label{fig_pattmW_all_norm}
\end{figure}

Fig.~\ref{fig_patt_all_norm} depicts the match rates of
\textit{Oracular} and \textit{OracularProj} normalized to NMP and NMP-Hyp, respectively. Each bar is marked by the NMP baseline used for comparison.
Overall, we observe that, both in near-term (\textit{Oracular}) and long-term (\textit{OracularProj}), \arch\ shows a significant improvement in throughput performance.
The maximum improvement is 133552$\times$ (for \textit{WC}) for long-term MTJ technology, due to good alignment of search and reference patterns in {\arch}. 
All applications have smaller improvement w.r.t. {NMP-Hyp}, both for near and long-term MTJ technologies, since  {NMP-Hyp} has no memory overhead and hence has a much higher match rate than {NMP} to start with. 
Fig.~\ref{fig_pattmW_all_norm} depicts the outcome for compute efficiency.
Generally we observe a similar trend to match rate, with all benchmarks (but {\em BC}) featuring >5$\times$ improvement even w.r.t. the ideal baseline {NMP-Hyp}.
Overall, {\em BC} shows the least benefit w.r.t. NMP-Hyp, 
since {\em BC} has a lower compute to memory access ratio and eliminating memory overhead greatly improves the {NMP-Hyp} throughput and compute efficiency.
\textit{RC4} has the highest improvements of approx. 300$\times$ and 900$\times$, for near-term and long-term respectively, in compute efficiency due to  
\arch's efficiency in handling its
high number of {XOR} operations.

\subsection{Gate-level Characterization}
\label{sec:ambit}
\begin{figure}[!]	
	\vshrink{0.2}
	\centering
	\includegraphics[width=0.37\textwidth]{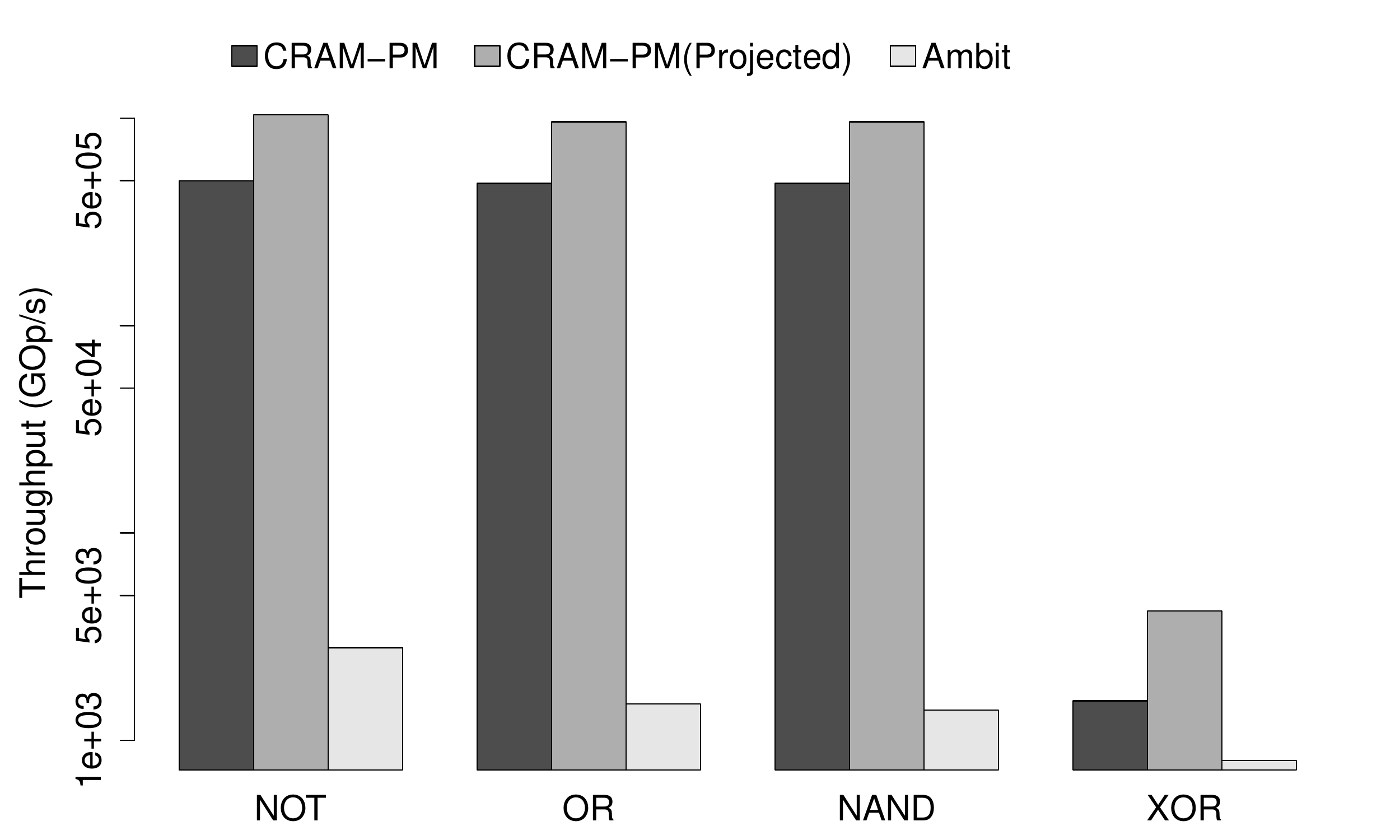}
	\vshrink{0.1}
	\caption{Throughput comparison w.r.t. Ambit~\cite{ambit}.}
	\label{fig_comparison}
	\vshrink{0.2}
\end{figure}

\noindent In this section, we compare
the throughput performance
of \arch\ with Ambit~\cite{ambit} and Pinatubo~\cite{li2016pinatubo}. Ambit reports a
comparative bulk throughput analysis with respect to CPU and GPU baselines, in
executing basic logic operations on fixed sized vectors of one-bit operands. Pinatubo reports bit-wise throughput of {OR} operation only, on a $2^{20}$ bit long vector. We considered the highest throughput (for 128-row operation) reported by Pinatubo. To conduct a fair comparison, we assume
the same vector
size of 32MB used in Ambit. 
\ignore{
Assuming a practical array size of $512 \times 512$, the entire vector 
would require 1368 arrays in total.
Each row in each tile stores $384$ one-bit operands and the results from bitwise
operations are stored in corresponding rows. 
To not favor \arch\ in this
analysis, we include all overheads while estimating SpinCM throughput i.e.
write, preset, read and bitline driver delays.  
}
Fig.\ref{fig_comparison} captures the outcome, w.r.t. Ambit,  in terms of Giga operations per
second (GOPs), for NOT, OR, NAND, and XOR implementations.
We observe a higher throughput for \arch\ across all of these bitwise
operations.
Ambit achieves the highest throughput for NOT, where \arch\ performs
approx. 178$\times$ and 370$\times$ 
better, considering near-term
and projected long-term MTJ technologies (Section~\ref{sec:eval_setup}),
respectively.  
The exploitation of row-level parallelism and
lack of actual data transfer within the array -- which is not the case for Ambit per
Section\ref{sec:rel} -- are the main reasons behind such improvement.
The
throughput of basic logic operations (i.e., NOT, OR, NAND) is very comparable to each other in \arch, unlike Ambit. 
For the more complex logic operation XOR, the
throughput improvement for long-term, projected \arch\ is 4$\times$ over Ambit; whereas for near-term
\arch,
only 1.34$\times$. In comparison to {OR} throughput of Pinatubo, \arch\ is approx. $6\times$ and $12\times$ better for near-term and long-term, respectively.   
For this comparison, we do not optimize data layout or operation scheduling for
\arch.
\ignore{
and the latency estimates are computed by an unoptimized design. Through proper
optimization of data layout and scheduling, the throughput difference will be
even higher. The compute and storage overhead was $\simeq10.6$ MB across all
tiles. This scratch-pad size can be reduced through layout and scheduling
optimization, and by trading off between throughput and the scratch-pad size.                 
}
That said, Ambit is based on a mature (DRAM) technology, and therefore more
versatile for integration in conventional systems.

\subsection{Impact of Process Variation}
\label{sec:process_var}

\section{Related Work}
\label{sec:rel}  
\noindent 
Without loss of generality, we base \arch\  on the spintronic PIM substrate CRAM which was briefly presented in~\cite{cram} and evaluated for a single-neuron digit recognizer along with a small scale 2D convolution in \cite{masoud}.
CRAM is unique in combining multi-grain (possibly dynamic) reconfigurability with true processing {\em in} memory semantics. The resistive {\em Associative Processor}~\cite{resistive_proc} and DRAM-based {\em DRAF}~\cite{DRAF} on the other hand, rely on look-up-tables to support reconfigurable fabrics like FPGA. 
The SRAM-based {\em Compute Cache}~\cite{computeCache} can carry out different vector operations
in the cache, but \arch\ 
needs a wider range of computations on much larger data than could fit in cache. 
Maintaining data coherence among cores which constitute near-memory logic is also an issue~\cite{lazypim,pnm_frame} which is not the case for \arch\ due
to the absence of dedicated cores (with full-fledged memory hierarchies) to implement logic operations.  

\arch\ performs true in-memory computation using STT-MRAMs. 
The idea is  configuring cells of the memory array as resistive dividers, since the state of an STT-MRAM cell corresponds to one of two resistance values. 
\ignore{
This idea does not directly extend to DRAMs, where the state corresponds to a charge, and the problems with performing similar
computations involve charge sharing or loss of state. 
DRAM-based solutions (e.g.,~\cite{michigan}) usually perform computations at the edge of memory. This implies that the bitlines required to transport data to the edge of memory constitute a serial bottleneck. \arch\ can simultaneously perform multiple operations in parallel in multiple rows. 
}
A comparable design based on memristors, MAGIC~\cite{magic}, also uses resistive division. 
Another work proposes an in-memory ReRAM based data parallel processor with SIMD ISA to implement complex functions for general purpose PIM~\cite{fujiki2018memory}.
Such arrays suffer from significant endurance issues when compared to STT-MRAMs. 
Recent proposals for bit-wise in memory computing include Ambit~\cite{ambit}, {Pinatubo}~\cite{li2016pinatubo} and STT-CiM~\cite{Jain}.   
Ambit~\cite{ambit} 
supports bitwise AND, OR, and NOT operations in DRAM, but
only performs computation on a designated set of rows.  Thus, to compute on an arbitrary row, the row must first be copied to these dedicated compute rows and then be copied back once the computation is complete. 
  {Pinatubo}~\cite{li2016pinatubo} on the other hand, can perform  bitwise operations on data residing in multiple rows, using a specialized sense amplifier with variable reference voltage, which increases the susceptibility to variation.
  \ignore{
  Bitwise
  operations are native to subarrays with MAT structures. When operations are
  inter-MAT and inter-bank, additional logic is required to accumulate
  the final result. Also, the parallelism utilized is constrained by the length of
  rows and allowed number of rows activated at the same time. 
  }
 %
  STT-CiM~\cite{Jain} is similar to Pinatubo, where multiple
  WL are activated to sense the logic function between data residing in
  participating rows. The difference is that STT-CiM supports more complex
  operations such as addition on top of basic Boolean functions. 
  The threshold
  current to sense amplifier is changed to achieve different logic
  functionalities. STT-CiM is also more susceptible to variation due to the use
  of sense amplifiers to execute logic functions.
  \ignore{
  Achievable parallelism is limited to the word length of
  STT-MRAM array. Also, since multiple columns are shared through the use of
  MUX, some bits in participating rows are not processed truly in parallel. It
  is also susceptible to variation in current through MTJ cells due to the use
  of sense amplifier to execute logic functions. Moreover, to accommodate vector
  operations, a dedicated reduction unit is employed in the design.}
  
  In~\cite{hypervec_associative_mem}, the functionality of human brain is imitated to solve pattern matching problems, where a learned hypervector is stored in a CAM structure and query patterns are matched one by one with the stored hypervector representation of reference pattern. While this approach might be suitable for approximate applications such as natural language processing, the inherent sequential nature of data processing limits the throughput. Also, the overhead of transforming data to a hypevector has a limiting contribution to the achievable throughput.   
  
FELIX~\cite{gupta2018felix} proposes a crossbar of memristors, which forms logic gates following the same principle as CRAM.
Although similar in concept, the majority and AND operations in FELIX are multi-cycle (vs. single cycle in \arch).
  Moreover, FELIX presents segmented bitlines (by inserting switches within bitlines) to make smaller arrays run in parallel and execute different operations on data. This approach can result in severe sneak current issues that can potentially prevent the design from functioning correctly.
  
  \ignore{
  Darwin ~\cite{darwin} is
  a framework which combines two features: D-SOFT, to filter search space and a
  hardware accelerated GACT, a constant memory based alignment algorithm, to
  perform alignment of arbitrarily long reads. While it provides a substantial
  speedup of over software implementation, it sacrifices accuracy. In an attempt
  to detect temporal correlations between event-based data streams using
  computational memory, PCM based memory array is
  used~\cite{sebastian2017temporal}. This work exploits the crystallization
  dynamics, on a system built around a PCM chip.
  }


\section{Conclusion}
\label{sec:conc}
\noindent This paper introduces \arch, a novel, reconfigurable spintronic
compute substrate for true in-memory pattern matching, which represents a key
computational step in large-scale data analytics.
When configured as memory, \arch\ is not any different than an MRAM array. Each
MRAM cell, however, can act as an input or output to a logic gate, on demand.
Therefore, reconfigurability does not compromise memory density. 
Each row can have only one logic gate active at a time, but the very same logic
operation can proceed in all rows (at the same columns) in parallel. 
We implement a proof-of-concept \arch\ array for large-scale character string
matching to pinpoint design bottlenecks and aspects subject to optimization. 
The
encouraging 
results from Section~\ref{sec:eval} indicate a great
potential for throughput performance and energy efficiency.

\ignore{
Based on the encouraging preliminary analysis from Section~\ref{sec:eval},
our future work is directed
to enable CRAM's energy effciency potential 
by bridging the
gap between the current and projected gains in energy efficiency.

\noindent In this paper, a novel processing-in-memory substrate, computational RAM or
CRAM, is proposed. This substrate provides true PIM features for in-place
computation. Based on arrays of spintronics device, MTJ, which are organized in
rows an columns of the array. Depending on the computational demand, the
individual devices can be connected to act as memory cells or logic cells, which
is configurable at runtime. Apart from energy and latency benefits from reducing
data transfer overhead to the minimum, it also leverages low energy and high
switching speed of MTJ technology. As a case study, we showed a basic DNA
sequence alignment application mapped to CRAM and analyzed the potential
performance benefits from such implementation. 
We observed potential at array granularity.... proof-of-cocenpt.

\subsection{Discussion} There has been a number of significant research efforts
to improve bitline driver design for STT-MRAM based designs.
In~\cite{bitline_hierarchial}, hierarchical bitline is proposed for eliminating
read disturbance using momentarily sharp MTJ cell current during the precession
period.  


}

\ignore{
This paper introduces Computational RAM, CRAM, a high-density reconfigurable
spintronics-based platform facilitating logic and memory. CRAM features true
processing-in-memory semantics, and by that differs from most CMOS-based
processing-near-memory solutions.

\ignore{
\begin{table}[htp]
\vshrink{0.1}
\caption{Potential technologies to implement CRAM.\label{tbl:tech}}
\vshrink{0.3}
\begin{center}
\resizebox{0.37\textwidth}{!}{	
  \begin{tabular}{ || c | c ||}
		\hline \hline
		{\bf Technology} &  {\bf State Variable} \\ \hline
		Voltage Controlled Logic & Resistance \\ \hline
		Memristor Logic & Resistance \\ \hline
		All-Spin Logic & Spin  \\ \hline
		(Classic) SRAM & Voltage \\ 
		\hline \hline
	\end{tabular}
  }
\end{center}
\end{table}
}

CRAM is unique in combining multi-grain (possibly dynamic) reconfigurability with true
processing {\em in} memory semantics.
The resistive element based 
{\em Associative Processor}~\cite{resistive_proc} and the DRAM technology based
{\em DRAF}~\cite{DRAF} on the other hand, both
represent look-up-table based solutions
which can support 
reconfigurable fabrics like FPGA.
A similar concept to CRAM based on resistive RAMs is introduced in
\cite{Strukov11}, but not a practical implementation.  The SRAM-based {\em
Compute Cache}~\cite{computeCache} 
can carry out different
vector operations such as copy, search, comparison in the cache, but CRAM can perform a wider range
of computations in the memory array.
Furthermore, mainitaining data coherence among cores which constitute PNM logic
is an issue~\cite{lazypim,pnm_frame} which is not the case for CRAM due to the absence of
dedicated cores (with full-fledged memory hierarchies) to implement logic
operations.  

This paper proposes a method for performing
true in-memory computation using STT-MRAMs. The CRAM design is based on
configuring segments of the memory array as resistive dividers, since the state of
an STT-MRAM cell is expressed as one of two resistance values. This idea does not
directly extend to DRAMs, where the state is stored as a charge, and the problems
with performing similar computations involve charge sharing, loss of state, etc.
In fact, to the best of our knowledge, there is no true in-memory approach available
for DRAM structures as prior methods (e.g.,~\cite{michigan}) perform
computations at the edge of memory. This implies that the bitlines required to
transport data to the edge of memory constitute a serial bottleneck. On the other
hand, in CRAM, it is possible to simultaneously perform multiple operations
in parallel in multiple rows.
A comparable design based on memristor-based technologies, MAGIC~\cite{magic},
also uses
resistive division. However, this work
does not go into a system-level evaluation of applications running on this
platform, as we do in this paper. At the same time, such
arrays suffer from significant endurance issues as compared to STT-MRAMs.

\ignore{
We have presented the CRAM concept in the context of an STT-MRAM
array, but the principles can also apply to other spin
or non-spin platforms (Table~\ref{tbl:tech}), including 
spin-Hall memories~\cite{Kim13},
or resistive 
RAMs (RRAMs)~\cite{Kawahara13},
with appropriate modifications
to the array, where nonvolatile solutions
render higher density arrays.
}

The CRAM architecture can perform computations locally within the array for any
intermediate operation, where major data movement tasks are generally 
only 
required for input data and the eventual output data.  In other words, over
an $n$-step operation where $n$ is sufficiently large, the cost of performing these
write operations can be amortized.  In contrast, in a conventional PNM
architecture, data must be taken out to the periphery of the memory array on
every operation, and then the result must be brought back to an array location,
incurring this type of overhead in each of the intermediate steps. 
Furthermore, the conventional PNM structure suffers from an inherent serial
bottleneck in that while one data set is being taken to the periphery, no other
operation may be performed within the array because the bit lines are shared
over all rows in the arrray. The CRAM array, on the other hand, can perform
operations in parallel {\em in every row} of the array because the logic lines
for each row are separate.

Based on the encouraging preliminary analysis from Section~\ref{sec:eval},
our future work is directed
to enable CRAM's energy effciency potential 
by bridging the
gap between the current and projected gains in energy efficiency.

\ignore{

\redHL{
While the CRAM concept can be implemented using
both spintronics- and CMOS-based technologies, when compared to CMOS-based,
spintronics-based solutions deliver better scalability, better memory array
density, and practically zero leakage power consumption. {\bf[R1.1]}
}

A similar concept to CRAM based on resistive RAMs is introduced in
\cite{Strukov11}, but not a practical implementation. In a recent attempt on
in-place processing, computation in cache is proposed which leverages emerging
bit-line technology in SRAM cache~\cite{computeCache}. This compute cache utilizes
large silicon area, previously dedicated to cache, to carry out different vector
operations such as copy, search, compare and logical operations.     
}

\ignore{

\begin{table}[htp]
\vshrink{0.1}
\caption{Potential technologies to implement CRAM.\label{tbl:tech}}
\vshrink{0.3}
\begin{center}
  \begin{tabular}{ | c | c | c | c |c |}
		\hline \hline
		$Technology$ & $State Variable$  & $Conversion$ & $Combined Variable$ & $Threshold Detector$\\ \hline
		$Voltage Controlled Logic$ & $Resistance$ & $V=IR$ & $Current$ & $MTJ$ \\ \hline
		$Memristor Logic$ & $Resistance$ & $V=IR$ & $Current$ & $Memristor$ \\ \hline
		$All-Spin Logic$ & $Spin$ & $-$ & $Spin$ & $Unstable Magnet$ \\ \hline
		$CMOS 6T SRAM$ & $Voltage$ & $I_{d}=K^{'}W/L[(V_{gs}-V_{t})V_{ds}-(V_{ds}^{2})/2]$ & $Current$ & $Inverter$ 
		\hline \hline
	\end{tabular}
\end{center}
\end{table}

We have presented the CRAM concept in the context of an STT-MRAM
array, but it is worth nothing that the principles can also apply to other spin
or non-spin platforms, including spin-Hall memories or resistive RAMs (RRAMs),
as long as the state variable corresponds to a resistance change.  

The CRAM platform facilitates true processing in-memory semantics, since each
cell can be configured to act as a logic or memory element to best match
application characteristics. The reconfigurability does not compromise memory
density. At the same time, by closely tracking the computational and memory
footprint of the workload, CRAM can minimize waste in both computational and
memory resources. 

Computing platforms with reconfigurability already exists in the literature for
a while now. Novel technologies e.g. spintronics based reconfigurable fabrics
are fairly well explored. There are some critical design parameters related to
these technologies which are the core obstacles to overcome in achieving
efficient designs, such as power consumption, scalability issue, lifetime etc.
In one research, resistive element based processor (Associative Processor) is
proposed, which is capable of near memory processing with very fine-grained
processing elements ~\cite{resistive_proc}. Although it presents attractive
solutions for implementation of reconfigurable fabrics like FPGA, it does not
have runtime reconfiguration capability. A significant amount of study is
attempted to address runtime reconfiguration issues in reconfigurable platforms.
A DRAM technology based solution (DRAF) is proposed by Gao $et. al.$ that
supports switching between multiple contexts with a very little timing overhead
~\cite{DRAF}. It separates itself from multi-context FPGA in the sense that it
does not rely on a backup off-chip memory to store the contexts, instead uses
dedicated on-chip high density DRAM cells. However, this solution also comes
with a few limitations. For example, the contents stored in LUT cannot be
changed during runtime, rather can only be accessed based on some runtime
condition. LUT contents and contexts are placed in the memory sub-arrays
statically (during synthesis) and cannot be reconfigured to reverse the role
during runtime. Moreover, only one of the contexts can be used at a time during
runtime, hence the support for parallel context execution is absent Such
constraints pose boundary on the application domains in which such
reconfigurable fabrics could be used. In another implementation with spintronics
devices, the high energy demand associated with conventional field-induced
magnetic switching based FPGA is addressed by Zhao $et. al.$ and STT-MRAM based
solution is proposed ~\cite{STTMRAM_FPGA}. This technique looks through runtime
configuration and energy issues, however the functionality of constituent blocks
are fixed during fabrication and hence is not \textit{truly} reconfigurable.      

}

}



\bibliographystyle{ieeetr}
\bibliography{cramdna}


\end{document}